\documentclass[a4paper,11pt]{article}
\pdfoutput=1 

\usepackage{jcappub} 
\usepackage{amssymb}
\usepackage{xcolor}
\usepackage[T1]{fontenc} 
\usepackage{amsmath}
\usepackage{amsfonts}
\usepackage{mathrsfs}
\usepackage[scr=rsfso,cal=zapfc,frak=euler,bb=ams]{mathalfa}
\usepackage{trimclip}
\usepackage{accents}

\DeclareUnicodeCharacter{2212}{-}

\title{Role of gravitational decoupling on theoretical insights of relativistic massive compact stars in the mass gap}

\author[a,1]{S. K. Maurya
\note{Corresponding author},}
\author[b]{A. Errehymy,}
\author[c]{Ksh. Newton Singh,}
\author[d]{G. Mustafa,}
\author[e]{Saibal Ray}

\affiliation[a]{Department of Mathematical and Physical Sciences,
College of Arts and Sciences, University of Nizwa, P.O. Box 33, Nizwa 616, Sultanate of Oman}
\affiliation[b]{Astrophysics Research Centre, School of Mathematics, Statistics and Computer Science, University of KwaZulu-Natal, Private Bag X54001, Durban 4000, South Africa}
\affiliation[e]{Department of Physics, National Defence Academy, Khadakwasla, Pune 411023, India}
\affiliation[d]{Department of Physics, Zhejiang Normal University, Jinhua 321004, China} 

\affiliation[e]{Centre for Cosmology, Astrophysics and Space Science (CCASS), GLA University, Mathura 281406, Uttar Pradesh, India}

\emailAdd{sunil@unizwa.edu.om}
\emailAdd{abdelghani.errehymy@gmail.com}
\emailAdd{ntnphy@gmail.com}
\emailAdd{gmustafa3828@gmail.com}
\emailAdd{saibal.ray@gla.ac.in}

\abstract{ Advancements in theoretical simulations of mass gap objects, particularly those resulting from neutron star mergers and massive pulsars, play a crucial role in addressing the challenges of measuring neutron star radii. In the light of this, we have conducted a comprehensive investigation of compact objects, revealing that while the distribution of black hole masses varies based on formation mechanisms, they frequently cluster around specific values. For instance, the masses observed in GW190814 $\left(23.2^{+1.1}_{-1.0} \, M_{\odot}\right)$ and GW200210 $\left(24.1^{+7.5}_{-4.6} \, M_{\odot}\right)$ exemplify this clustering. We employed the gravitational decoupling approach within the framework of standard general relativity and thus focusing on the strange star model. This model highlights the effects of deformation adjusted by the decoupling constant and the bag function. By analyzing the mass-radius limits of mass gap objects from neutron star mergers and massive pulsars, we can effectively constrain the free parameters in our model, allowing us to predict the radii and moments of inertia for these objects. The mass-radius ($M-R$) and mass-inertia ($M-I$) profiles demonstrate the robustness of our models. It is shown that as the decoupling constant $\beta$ increases from 0 to 0.1 and the bag constant $\mathcal{B}_g$ decreases from 70 $\text{MeV/fm}^3$ to 55 $\text{MeV/fm}^3$, the maximum mass reaches $M_{\text{max}} = 2.87 \, M_\odot$ with a radius of 11.20 km. In contrast, for $\beta = 0$, the maximum mass is $M_{\text{max}} = 2.48 \, M_\odot$ with a radius of 10.69 km. Similarly, it has been exhibited that as $\beta$ decreases to 0, the maximum mass peaks at $M_{\text{max}} = 2.95 \, M_\odot$ for $\mathcal{B}_g = 55 \, \text{MeV/fm}^3$ with a radius of 11.32 km. These results not only exceed the observed masses of compact stars but also correlate with recent findings from gravitational wave events like GW190814 and GW200210, underscoring the relevance of our models in exploring compact objects in the universe.} 

\begin{document}

\maketitle

\flushbottom

\section{Introduction }
The detection of gravitational waves by the LIGO Scientific and VIRGO Collaboration has significantly advanced our understanding of classical general relativity (GR), particularly regarding its limitations in modeling supermassive black holes (BHs). Notable events such as GW190814 and GW170817 have provided critical insight into these deficiencies \cite{LIGOScientific:2016aoc,LIGOScientific:2017ync}. The GW190814 event highlights a merger between a compact binary system consisting of a BH, estimated to have a mass between 22.2 and 24.3 $ M_{\odot}$, and a companion object with a mass ranging from 2.50 to 2.67 $ M_{\odot}$. Meanwhile, the GW170817 event, detected on August 17, 2017, is thought to be the merger of two NSs, each with masses between 0.86 and 2.26 $ M_{\odot}$. These gravitational wave events are particularly significant in the context of pulsars (PSs), strange stars (SSs), and neutron stars (NSs), which have long served as cosmic laboratories for studying matter at ultrahigh densities. Although Einstein's classical theory of GR accounts for various observed phenomena, such as compactness, mass-radius relationships, and surface redshifts, it fails to explain the unusual observations associated with NSs that exceed a mass of $M=2 M_{\odot}$. This discrepancy underscores the importance of ongoing research into the nature of these compact objects and the fundamental laws of physics that govern them. 

Several theories have been proposed to explain the observed signals from gravitational events. These include the characteristics of matter via equation of state (EOS) in the binary star system and various modifications to gravity theories. The signal detected from GW190814 has sparked discussions on the nature of the secondary component, influenced by several factors, including its mass of approximately $2.59^{+0.08}_{-0.09} M_{\odot}$, the absence of significant tidal deformations, and the lack of accompanying electromagnetic signals. This indicates the possibility that the secondary component is an NS or a BH. Consequently, we face a dilemma: it might be the lightest BH or the heaviest NS ever observed in a double compact object system. In the meantime, the GW170817 event and its associated electromagnetic signals offered scientists a valuable opportunity to suggest more unconventional EOSs for the NSs involved in this merger. The electromagnetic radiation associated with GW170817 included two main features: a brief gamma-ray burst, GRB170817A, that was detected approximately 2 seconds after the gravitational wave signal, and a kilonova referred to as AT2017gfo, which peaked in brightness several days after the merger. 

Several researchers have been investigating the properties of the binary components based on the observed delays. The kilonova signal indicates a relatively stiff EOS \cite{Guo:2023vzz, Radice:2017lry}. To explain the potential tidal deformability range of 400 to 800 and the radius of the 1.5 solar mass component, estimated between 11.8 km and 13.1 km -- the GW170817 event has been modeled as a merger between a hadronic star (HS) and a SS \cite{Burgio:2018yix}. Furthermore, the delay in the gamma-ray burst has led some researchers to suggest that the remnant of the merger was probably a hypermassive star that collapsed into a BH in just a few milliseconds \cite{Ruiz:2017due}. To address the observed small radii and moderate tidal deformability, it has also been proposed that stellar matter undergoes significant phase transitions at supranuclear densities, resulting in the formation of quark matter. In addition, the recently announced GWTC-3 catalogue by the LVC includes new asymmetric systems \cite{KAGRA:2021duu}. Among these, GW200210 stands out as a compact-object merger, showing notable similarities to GW190814. Despite being a relatively low-significance detection (false-alarm rate $> 1~yr^{−1}$), its discovery suggests that asymmetric systems may be more common than previously thought.

Observations of galactic compact objects suggest a shortage of binaries with masses in the lower mass gap \cite{Bailyn:1997xt,Ozel:2010su,Farr:2010tu,ArcaSedda:2021zmm,Bhar2023}, potentially linked to star formation processes, although selection effects may also play a role. The kicks imparted by core collapse supernovae might lead to higher velocities for lower-mass BHs, making them less likely to stay in observable binaries. Compact objects can reach mass within this gap through gas accretion -- either from low-mass X-ray binaries or AGN disks -- or via mergers, such as NS collisions in dense stellar clusters. However, the expected rates of these mergers are low, especially for the masses observed \cite{Samsing:2019dtb, Ye:2019xvf, Fragione:2020aki, Lu:2020gfh}. BHs in cluster cores tend to dominate, limiting NS mass segregation and leading to a higher rate of BH mergers compared to those involving NSs. Consequently, the likelihood of dynamical origins or AGN-assisted binary mergers involving objects in the lower mass gap appears minimal, as most mergers are expected to be BH-BH events. NS mergers may occur within hierarchical triple systems, where two NSs form from an inner binary and a BH from a tertiary star \cite{Lu:2020gfh, Liu:2020gif}. The expected rate of such double mergers is about 0.1\% to 3\% of the binary NS merger rate of 10--1700 $Gpc^{-3} yr^{-1}$, suggesting that some observed NS mergers could lead to secondary mergers \cite{LIGOScientific:2020zkf, Bartos:2023lfu}.

The origins of objects in the lower mass gap can be inferred from their mass. Known Galactic binary NS systems display a narrow mass distribution centered around approximately $2.65 \pm 0.12 \, M_{\odot}$ \cite{Farrow:2019xnc}, suggesting that the remnants of NS mergers likely cluster around a mass similar to about $2.6 \, M_{\odot}$ \cite{Lu:2020gfh}. In contrast, processes such as accretion and stellar evolution may result in a broader mass distribution in this gap, as NSs must accrete several tenths of a solar mass to reach these levels. Even if accretion leads to a narrower distribution, there is no definitive astrophysical reason for it to center around $2.6 \, M_{\odot}$. Additionally, if two NSs become gravitationally bound through chance encounters in stellar clusters, they may not originate from a binary system, leading to a wider mass distribution compared to binary NSs \cite{Kiziltan:2013oja, Valentim:2011vs, Zhang:2010qr}. For mass gap objects formed by NS mergers, the BH mass distribution may vary depending on their origin. For instance, GW190814 $(23.2^{+1.1}_{-1.0} \, M_{\odot})$ and GW200210 $(24.1^{+7.5}_{-4.6} \, M_{\odot})$ have similar primary masses. In stellar-cluster scenarios, merger remnants encounter BHs randomly, producing a broader primary-mass distribution. Conversely, hierarchical triple scenarios may lead to a more narrowly distributed primary mass, especially when a heavier tertiary BH stabilizes the system \cite{Lu:2020gfh}, as supernova explosions are less disruptive in these configurations, allowing outer orbits to remain stable.

Our main aim is to simulate the configurations of matter associated with mass gap objects formed by NS mergers and other massive pulsars. While the distribution of BH masses can vary depending on their formation pathways, they often cluster around similar values, such as those seen in GW190814 $(23.2^{+1.1}_{-1.0} \, M_{\odot})$ and GW200210 $(24.1^{+7.5}_{-4.6} \, M_{\odot})$. We shall investigate these phenomena using the gravitational decoupling (GD) approach \cite{Ovalle:2017fgl, Ovalle:2018gic}, with additional insights provided in \cite{daRocha:2020jdj, Ovalle:2017wqi, Ovalle:2018umz, Ovalle:2019lbs, Casadio:2019usg, Zubair:2020lna, Ovalle:2020kpd, Ovalle:2021jzf, Contreras:2021yxe, Maurya:2024ylr, Maurya:2022cyv} that explore the applications of GD within the framework of standard GR. For further applications in higher dimensions, see \cite{Maurya:2022uqu, Maurya:2022brt}, for instance. In our examination of the SS model, we place special emphasis on the contributions from the deformation adjusted by the decoupling constant and the special bag function within the bag model. This framework incorporates all necessary adjustments to the energy and pressure functions of the SQM approach, which posits that quarks are noninteracting and massless. By analyzing the observed mass-radius limits of mass-gap objects formed by NS mergers and other massive pulsars, we can effectively constrain the free parameters in our model. This, in turn, enables us to make predictions about the radii and moments of inertia of these objects.

This work is organized as follows: In Sect. \ref{sec2}, we present the gravitationally decoupled Einstein field equations produced by two sources. Sect. \ref{sec3} explores the minimally deformed solution for SSs, with Subsect. \ref{solA} addressing the density constraint, $\epsilon(r) = \theta^0_0(r)$, and Subsect. \ref{solA} focusing on the pressure constraint, $P_r(r) = \theta^1_1(r)$. In Sect. \ref{sec4}, we discuss the boundary conditions for SS models under gravitational decoupling. In Sect. \ref{sec5}, we will explore the physical viability of our deformed strange star models and their implications in astrophysics. The stability analyses, including those based on the adiabatic index and the Harrison-Zel'dovich-Novikov criteria, are discussed in Sect. \ref{sec6}. Finally, concluding remarks can be found in Sect. \ref{sec7}.

\section{Gravitationally decoupled Einstein field equation (EFE) generated by two sources }\label{sec2}
This section provides a focused overview of gravitationally decoupled Einstein field equations for two independent sources
\begin{eqnarray} \label{eq1}
 &&   G_{ij}=R_{ij}-\frac{1}{2}\,g_{ij}\,\mathcal{R}= -8\pi T^{\text{eff}}_{ij}
\end{eqnarray}
with
\begin{eqnarray}\label{eq2}
T^{\text{eff}}_{ij} =T_{ij}+\beta\,\theta_{ij},
\end{eqnarray}
where the relativistic units are denoted as $G = c = 1$.

The Ricci tensor is designated by the notation $R_{ij}$, where $\mathcal{R}$ represents the contracted Ricci scalars and $\beta$ denotes the decoupling parameter to the field equations. The energy-momentum tensor is denoted as $T_{ij}$, and the source $\theta_{ij}$ has the potential to define other fields, including scalar, vector, and tensor fields. The fact that the Einstein tensor $(G_{ij})$ satisfies the Bianchi identity implies that the effective energy-momentum tensor $T^{\text{eff}}_{ij}$ must be preserved
\begin{eqnarray}\label{eq3}
&& {\nabla_i}[\,T^{ij}\,]^{\text{eff}}=0.
\end{eqnarray}

The following static spherically symmetric line element is being utilized to describe the space-time of the star system's internal region:
\begin{eqnarray} \label{eq4}
 ds^2 = -e^ {\mathcal{N}_0 (r)} dr^2 - r^2\big(d\theta^2 + \sin^2 \theta~ d\phi^2\big)~+ e^{\mathcal{F}_0( r)} dt^2 ,
 \end{eqnarray}
where the metric potentials $\mathcal{F}_0$ and $\mathcal{N}_0$ are the only ones that are dependent on radial distance. 

Taking into consideration the fact that the internal structure of the self-gravitating system, which is characterized by the effective energy-momentum tensor $T^{\text{eff}}_{ij}$, describes an anisotropic matter distribution
\begin{eqnarray}
&& T^{\text{eff}}_{ij} = \left({\rho^{\text{eff}}}+{P^{\text{eff}}_t}\right)u_{i}u_{j}-{P^{\text{eff}}_t} g_{ij}+({P^{\text{eff}}_r}-{P^{\text{eff}}_t})\chi_{i}\,\chi_{j},\label{eq5}
\end{eqnarray}
where the formula for $u^i$ is defined as: $u^i=e^{\mathcal{F}_0(r)/2}{\delta^i}_4$, therefore denoting the four-velocity. The unit vector in the radial direction is denoted by the symbol $\chi^i$, which is defined as $\chi^i = e^{\mathcal{N}_0(r)/2} {\delta^i}_1$.
The pressures in the radial and tangential directions are represented by $P^{\text{eff}}_r$ and $P^{\text{eff}}_t$, respectively, while the energy density of matter is characterized by $\rho^{\text{eff}}$. Furthermore, the $4$-velocity $\chi^{i}$ and the unit space-like vector $u^{i}$ in the radial direction satisfied the conditions: $\chi^{}$.  Therefore, the components of the effective energy-momentum tensor $T^{\text{eff}}_{ij}$ can be provided as follows:
\begin{eqnarray}\label{eq6}
[T^0_0]^{\text{eff}}=\rho^{\text{eff}}, ~~[T^1_1]^{\text{eff}}=-P^{\text{eff}}_{r},~~\text{and}~~[T^2_2]^{\text{eff}}=-P^{\text{eff}}_t.~~~
\end{eqnarray}

Next, the complete formulation of Einstein's field equations may be expressed as the collection of the following differential equations for the metric~(\ref{eq1}):
\begin{eqnarray}
-\kappa [T^1_1]^{\text{eff}}&=& \frac{1}{8\pi}\bigg[-\frac{1}{r^2}+e^{-\mathcal{N}_0}\left(\frac{1}{r^2}+\frac{\mathcal{F}_0^{\prime}}{r}\right)\bigg]=\kappa\, P^{\text{eff}}_r ,\label{eq9}\\
 -\kappa [T^2_2]^{\text{eff}}= -\kappa [T^3_3]^{\text{eff}} &=& \frac{1}{8\pi}\bigg[\frac{e^{-\mathcal{N}_0}}{4}\left(2\mathcal{F}_0^{\prime\prime}+\mathcal{F}_0^{\prime2}-\mathcal{N}_0^{\prime}\mathcal{F}_0^{\prime}+2\frac{\mathcal{F}_0^{\prime}-\mathcal{N}_0^{\prime}}{r}\right)\bigg]= \kappa\, P^{\text{eff}}_t ,~~\label{eq10} \\
\kappa [T^4_4]^{\text{eff}} &=&  \frac{1}{8\pi}\bigg[\frac{1}{r^2}-e^{-\mathcal{N}_0}\left(\frac{1}{r^2}-\frac{\mathcal{N}_0^{\prime}}{r}\right)\bigg]= \kappa \rho^{\text{eff}}.\label{eq11}
\end{eqnarray}

Applying the metric function allows for the determination of the mass function, $m(r)$, of a charged perfect fluid sphere
\begin{equation}
e^{-\mathcal{N}_0} = 1 - \frac{2m(r)}{r},\label{eq12}
\end{equation}
one can get also as equivalent to~\cite{Sharp-Misner1964} 
\begin{equation}
m(r)= \frac{\kappa}{2} \int \rho^{\text{eff}}\, r^2 dr.\label{eq14}
\end{equation}

By using Eqs. (\ref{eq9}) and (\ref{eq12}), we derive
\begin{equation}
\frac{\mathcal{F}_0^\prime }{2}= \frac{\kappa r P^{\text{eff}}_r + \frac{2m}{r^2} }{1 - \frac{2m}{r} }.\label{eq15}
\end{equation}

On the assumption that $P^{\text{eff}}_r \neq P^{\text{eff}}_t$, the condition $P^{\text{eff}}_r = P^{\text{eff}}_t$ suggests the presence of an isotropic fluid distribution. The anisotropy factor, represented by the symbol $\Delta$, is precisely mathematically defined as: $\Delta^{\text{eff}}=P^{\text{eff}}_t-P^{\text{eff}}_r$. The term $\frac{2(P^{\text{eff}}_t-P^{\text{eff}}_r)}{r}$ denotes a force that emerges from the anisotropic properties of the fluid. When the pressure $P^{\text{eff}}_t$ exceeds the pressure $P^{\text{eff}}_r$, the force is directed in an outward direction. When the value of $P^{\text{eff}}_t$ is less than $P^{\text{eff}}_r$, the force acts with an inward direction. Nevertheless, if the value of $P^{\text{eff}}_t$ is greater than the value of $P^{\text{eff}}_r$, the force allows for the formation of a more condensed structure under the scenario of an anisotropic fluid as opposed to isotropic fluid distribution~\cite{Gokhroo1994}.

In addition to Eqs. (\ref{eq12}) and (\ref{eq15}), the pressure gradient may also be represented in terms of $m$, $q$, $\rho$, and $P_r$ using equations (\ref{eq9})--(\ref{eq11})
\begin{equation}
\frac{dP^{\text{eff}}_r}{dr} =- \frac{\kappa r P^{\text{eff}}_r + \frac{2m}{r^2}}{1 - \frac{2m}{r} } (P^{\text{eff}}_r+\rho^{\text{eff}})+ \frac{2 (P^{\text{eff}}_t-P^{\text{eff}}_r)}{r},\label{eq7}
\end{equation}
which provides the generalized Hydrostatic TOV equation for anisotropic stellar structure~\cite{TOV1,TOV2}.  

Our next approach is to determine a precise solution for the field equations (\ref{eq9})--(\ref{eq11}) for the TOV equation (\ref{eq7}) that describes a model of a strange star. We see that the system of field equations exhibits a significant degree of non-linearity, thereby presenting a challenging task in their solution.  Hence, we use an alternative method called gravitational decoupling using the minimum geometric deformation (MGD) methodology and a specific transformation associated with the gravitational potential
\begin{eqnarray}
&& \mathcal{F}_0(r) \longrightarrow \mathcal{A}_0(r)+\beta\, \mathcal{H}(r), \label{eq26}\\
&& e^{-\mathcal{N}_0(r)} \longrightarrow \mathcal{B}_0(r)+\beta\, \mathcal{G}(r).  \label{eq27}
\end{eqnarray}

Let $\mathcal{H}(r)$ and $\mathcal{G}(r)$ represent the decoupling functions, respectively, in connection with the temporal and radial metric components. The deformation may be appropriately tuned by adjusting the decoupling constant $\beta$. When $\beta =0$, the conventional GR-gravity theory is systematically restored. By using the MGD technique, we are able to establish the values of $\mathcal{H}(r)=0$ and $\mathcal{G}(r)\ne0$. This observation suggests that the appropriate transformation only applies to the radial component of the metric function, while the temporal component remains unaltered. This method of MGD partitions the decoupled system (\ref{eq9})--(\ref{eq11}) into two distinct parts. The initial system is associated with $T_{ij}$, whereas the second system relates to the additional source $\theta_{ij}$. To formulate the initial system, we examine the energy-momentum tensor $T_{ij}$ which characterizes an anisotropic matter distribution presented by
\begin{equation}\label{eq28}
T_{ij}=\left(\epsilon+P_t\right)u_{i}\,u_{j}-P_t\,\delta_{ij}+\left(P_r-P_t\right)\chi_{i}\,\chi_{j}.
\end{equation}

The energy density is denoted by $\epsilon$, while the radial pressure and tangential pressure for the seed solution are denoted by $p_{r}$ and $p_{t}$ correspondingly. Accordingly, the effective quantities may be expressed as
\begin{eqnarray}
\rho^{\text{eff}}=\epsilon+\beta \,\theta^0_0,~~P^{\text{eff}}_r=P_r-\beta\,\theta^1_1,~~P^{\text{eff}}_t=P_t-\beta\,\theta^2_2.~~~~  \label{eq29}
\end{eqnarray}

Furthermore, the related effective anisotropy is
\begin{eqnarray} 
&&\hspace{-0.7cm} \Delta^{\text{eff}}=P^{\text{eff}}_t-P^{\text{eff}}_r= \Delta_{GR}+\Delta_{\theta},
 \label{eq30}\\
&&\hspace{-0.7cm} \text{where}~~~~\Delta_{GR}= P_t-P_r~~~~~\text{and}~~~~~\Delta_\theta= \beta (\theta^1_1-\theta^2_2)\nonumber. 
\end{eqnarray}

The effective anisotropy can be defined as the combined value of two anisotropies in connection to matter distribution namely $T_{ij}$ and $\theta_{ij}$.  Gravitational decoupling generates the anisotropy ($\Delta_{\theta}$), which may increase the effective anisotropy. The system (\ref{eq9})--(\ref{eq11}) may be decomposed into two systems by employing the transformations (\ref{eq26}) and (\ref{eq27}). The first system is dependent on the gravitational potentials $\mathcal{A}_0$ and $\mathcal{B}_0$, (viz. when $\beta = 0$):
\begin{eqnarray}
&&\hspace{-0.8cm}\epsilon= \frac{1}{8\pi} \bigg(\frac{1 }{r^2}-\frac{\mathcal{B}_0  }{r^2}-\frac{\mathcal{B}_0^{\prime} }{r} \bigg),\label{eq19}\\
&&\hspace{-0.8cm} P_r=\frac{1}{8\pi} \bigg(-\frac{1 }{r^2}+\frac{\mathcal{B}_0  }{r^2}+\frac{\mathcal{A}_0^{\prime} \mathcal{B}_0  }{r}\bigg), \label{eq20}\\
&&\hspace{-0.8cm} P_t=\frac{1}{8\pi} \bigg(\frac{\mathcal{B}_0^{\prime} \mathcal{A}_0^{\prime}  }{4}+\frac{\mathcal{A}_0^{\prime \prime} \mathcal{B}_0  }{2}+\frac{\mathcal{A}_0^{\prime 2} \mathcal{B}_0 }{4}  +\frac{\mathcal{B}_0^{\prime}  }{2 r}+\frac{\mathcal{A}_0^{\prime} \mathcal{B}_0  }{2 r}\bigg). \label{eq21}
\end{eqnarray}

From Eq. (\ref{eq7}), the following result is obtained:
\begin{eqnarray}
-\frac{\mathcal{A}_0^\prime}{2}(\epsilon+P_r)-P_r^{\prime}+\frac{2}{r}( P_{t}-P_r)=0.~~\label{eq22}
\end{eqnarray}

This is a TOV equation for the system's configuration (\ref{eq19})--(\ref{eq21}), a solution to which may be found in the spacetime presented below:
\begin{equation}\label{eq35}
ds^2=e^{\mathcal{A}_0(r)}dt^2-\frac{dr^2}{\mathcal{B}_0(r)}-r^2d\theta^2+r^2\text{sin}^2\theta \,d\phi^2.
\end{equation}

By activating $\beta$, one may derive the second set of equations as
\begin{eqnarray}
&&\hspace{-0.7cm}\theta^{0}_0=-\frac{1}{8\pi}\Big(\frac{\mathcal{G}   }{r^2}+\frac{\mathcal{G}^\prime }{r}\Big), \label{eq36}\\
&&\hspace{-0.7cm}\theta^1_1=-\frac{1}{8\pi}\Big(\frac{\mathcal{G}  }{r^2}+\frac{\mathcal{A}_0^{\prime} \mathcal{G}   }{r}\Big), \label{eq37}\\
&&\hspace{-0.7cm}\theta^2_2=- \frac{1}{8\pi}\Big(\frac{1}{4} \mathcal{G}^\prime \mathcal{A}_0^{\prime}   +\frac{1}{2} \mathcal{A}_0^{\prime \prime} \mathcal{G}   +\frac{1}{4} \mathcal{A}_0^{\prime 2} \mathcal{G}  +\frac{\mathcal{G}^\prime  }{2 r}+\frac{\mathcal{A}_0^{\prime} \mathcal{G}  }{2 r}\Big). \label{eq38}
\end{eqnarray}

The following resulting equation is provided by the linear combination of the Eqs. (\ref{eq36})--(\ref{eq38}) as
\begin{eqnarray}
-\frac{\mathcal{A}_0^{\prime}}{2} (\theta^0_0-\theta^1_1)+ (\theta^1_1)^\prime+\frac{2}{r} ~(\theta^1_1-\theta^2_2)=0. \label{eq39}
\end{eqnarray}

The mass distribution for each system may be expressed by the following formula
\begin{eqnarray}
&&\hspace{-0.7cm} m_{Q}=\frac{1}{2} \int^r_0 \rho(x)\, x^2 dx~~~\text{and}~~~m_{\theta}= \frac{1}{2}\,\int_0^r \theta^0_0 (x)\, x^2 dx, \label{eq40}
\end{eqnarray}
where the mass functions corresponding to sources $T_{ij}$ and $\theta_{ij}$ are indicated by the variables $m_{GR}(r)$ and $m_{\theta}(r)$, respectively. 
 
Finally, the benefit of MGD-decoupling gets clear: we can generalize any existing solutions that are linked to the matter-spacetime $\{T_{ ij}, \mathcal{A}_0, \mathcal{B}_0\}$ provided by Eqs. (\ref{eq9})--(\ref{eq11}) and by solving the unconventional gravitational system of equations Eqs. (\ref{eq36})--(\ref{eq38}) to find $\{\theta_{ ij }$, $\mathcal{H}$, $\mathcal{G}\}$. Therefore, we may produce the "$\theta$-version" of any $\{T_{ij} \mathcal{A}_0, \mathcal{B}_0\}$-solution as 
\begin{eqnarray}
\{T_{ ij},~ \mathcal{A}_0(r),~ \mathcal{B}_0(r)\} \Longrightarrow \{T^{\text{eff}}_{ ij}, ~\mathcal{F}_0(r),~~\mathcal{N}_0(r)\}.
\end{eqnarray}

The above connection outlines a direct approach to investigate the effects of gravity that extend beyond standard Einstein's gravity.

\section{Minimally deformed solution for strange star (SS) } \label{sec3}

In order to get a non-singular, monotonically decreasing matter density within the spherically symmetric star system, we choose a modified version of $\epsilon$ as proposed by Mak and Harko \cite{Harko:2002pxr}:
 \begin{eqnarray}
     \epsilon(r)=\epsilon_0 \bigg[ 1-\bigg(1-\frac{\epsilon_s}{\epsilon_0}\bigg) \frac{r^2}{r^2_s} \bigg]. \label{eq2.1}
 \end{eqnarray}

The constants $\epsilon_0$ and $\epsilon_s$ represent the central and surface maximum and minimum values of $\epsilon$, respectively. For the sake of studying the SS model, we make the assumption that the distribution of SQM within the peculiar stars is controlled by the straightforward physical MIT bag model EOS \cite{Chodos:1974}. The introduction of the special bag function within the bag model has preserved all the adjustments of the energy and pressure functions of the SQM approach. Our simple bag model assumes that the quarks are both non-interacting and massless. The quark pressure shall be defined as  
\begin{eqnarray} \label{eq41}
P_r=\sum_{f=u,~d,~s} P^{f}-\mathcal{B}_g.
\end{eqnarray}

Let $P^f$ represent the individual pressures associated with the quark flavors $(u)$, $(d)$, and $(s)$. These pressures are neutralized by the total external Bag pressure or Bag constant $\mathcal{B}_g$. The energy density ($\epsilon$) for the deconfined quarks interior, in relation to the MIT Bag model, can be expressed as following: 
\begin{eqnarray}
\epsilon=\sum_{f}\epsilon^{f}+\mathcal{B}_g,~~~\text{where}~~~\epsilon^f=3P^f.\label{eq42}
\end{eqnarray}

Applying Eqs. (\ref{eq41}) and (\ref{eq42}) along with the connection $\epsilon^f=3P^f$, we may express the MIT bag EOS for strange quark stellar objects in its explicit form
\begin{eqnarray} \label{eq2.4}
P_r=\frac{1}{3}(\epsilon-4\mathcal{B}_g).
\end{eqnarray} 

At this stage, we have clearly defined the realistic matter density and realistic EOS. Now we focus on the spacetime geometries $\mathcal{A}_0$ and $\mathcal{B}_0$ for the first system. The differential equation derived from Eqs. (\ref{eq2.1}) and (\ref{eq19}) is expressed as follows:  
\begin{eqnarray}
r \mathcal{B}_0'(r)+\mathcal{B}_0(r)+\frac{r^4 (\epsilon_s-\epsilon_0)+r^2 \epsilon_0 r_s^2-8 \pi  r_s^2}{8 \pi  r_s^2}=0.
\end{eqnarray}

After doing the integration, we find the potential $\mathcal{B}_0$ as follows:
\begin{eqnarray}
 \mathcal{B}_0=\frac{24 \pi  r^4 (\epsilon_0-\epsilon_s)-40 \pi  \epsilon_0 r^2 r_s^2+15 r_s^2}{15 r_s^2}. \label{eq2.6}
\end{eqnarray}

To find the other potential, we combine the EOS (\ref{eq2.4}) and Eqs. (\ref{eq20}) and (\ref{eq2.1}), which yields subsequent differential equation
\begin{eqnarray}
5 r_s^2 (32 \pi  \mathcal{B}_g r+3 \mathcal{A}'_0)+8 \pi  r_0 r \left(3 \mathcal{A}'_0 r^3-5 \mathcal{A}'_0 r r_s^2+8 r^2-10 r_s^2\right)-8 \pi  r_s r^3 (3 \mathcal{A}'_0 r+8) =0. \label{eq2.7}
\end{eqnarray}

Integrating above Eq.~(2.7) that yields a following solution for $\mathcal{A}_0(r)$ as
\begin{eqnarray}
    \mathcal{A}_0(r) &=&\frac{2 \sqrt{10 \pi } r_s (\epsilon_0-6  \mathcal{B}_g) }{3 \sqrt{10 \pi  \epsilon_0^2 r_s^2-9 \epsilon_0+9 \epsilon_s}}\tanh ^{-1}\left(\frac{\sqrt{\frac{2 \pi }{5}} \left(-6 \epsilon_0 r^2+5 \epsilon_0 r_s^2+6 \epsilon_s r^2\right)}{r_s \sqrt{10 \pi  \epsilon_0^2 r_s^2-9 \epsilon_0+9 \epsilon_s}}\right)\nonumber\\&& -\frac{2}{3} \log \left(8 \pi  \epsilon_0 r^2 \left(3 r^2-5 r_s^2\right)-24 \pi  \epsilon_s r^4+15 r_s^2\right)+\mathcal{C}.  \label{eq2.8}
\end{eqnarray}

This set of equations (\ref{eq2.6}) and (\ref{eq2.8}) provides the complete spacetime geometry for the seed solution. Still, given the $\theta$-sector, it is necessary to determine the result of the second set of equations (\ref{eq36})--(\ref{eq38}). Nevertheless, it is important to acknowledge that there are three autonomous equations using four unknown variables. Hence, this scenario necessitates the inclusion of one more piece of information to complete the $\theta$-system, such as $\mathcal{G}(r)$. Considering this problem, we choose the consequent choices:
\begin{eqnarray}
&&\hspace{-0.5cm} \mathcal{G}(r)=\mathcal{B}_0(r)-1,~~\label{eq58}\\
&&\hspace{-0.5cm} \mathcal{G}(r)=\frac{1}{ 1+r\,\mathcal{A}_0'(r)}- \mathcal{B}_0 (r).~~~~~~~\label{eq59}
\end{eqnarray}

It can be seen that $\mathcal{G}(r)$ is devoid of any singularities and that $\mathcal{G}(0)$ is equal to zero. Consequently, they enable us to simulate (i) the component $\theta^0_0$ with the energy density $\epsilon$, namely $\epsilon=\theta^0_0$ as defined in Eq.~(\ref{eq58}), and (ii) the radial pressure $P_r$ with the component $\theta^1_1$ as defined in Eq.~(\ref{eq59}). This approach has been effectively used by many authors \cite{Ovalle2017,OvalleEPJC2018,sharif2,Maurya:2023muz,Maurya:2021zvb} to describe compact objects in GR and alternative gravity theories.  

Inspired by those studies, we apply the mimic technique in our current work to solve the system of equations (\ref{eq39})--(\ref{eq41}). Thus, we use the following methodologies $\rho=\theta^0_0$ and $p_r=\theta^1_1$ to investigate the effects of two density constraints in this study.

\subsection{Mimicking of the Density Constraint: $\epsilon(r)=\theta^0_0(r)$} \label{solA}

By mimicking of the seed density ($\epsilon$) to component ($\theta^0_0$) through the equations (\ref{eq19}) and (\ref{eq36}), we derive  
\begin{eqnarray}
  r \frac{d\mathcal{G}}{dr}  + \mathcal{G}   -\frac{8\pi r^2\left[r^2 (\epsilon_0-\epsilon_s)-\epsilon_0 r_s^2\right]}{r_s^2}=0.
\end{eqnarray}

After solving the above equation, we derive the exact solution of decoupling function $\mathcal{G}(r)$ as
\begin{eqnarray}
    \mathcal{G}(r)=\frac{24 \pi  r^4 (\epsilon_0-\epsilon_s)-40 \pi  \epsilon_0 r^2 r_s^2}{15 r_s^2}.\label{eq2.12} 
\end{eqnarray}

Then, the deformed MGD solution for the density constraint approach can be expressed by the following spacetime geometries
\begin{eqnarray}
 && \hspace{-0.5cm}   e^{\mathcal{F}_0(r)}= e^{\mathcal{F}_0(r)}= \exp\Bigg[\frac{2 \sqrt{10 \pi } r_s (\epsilon_0-6  \mathcal{B}_g) }{3 \sqrt{10 \pi  \epsilon_0^2 r_s^2-9 \epsilon_0+9 \epsilon_s}}\tanh ^{-1}\left(\frac{\sqrt{\frac{2 \pi }{5}} \left(-6 \epsilon_0 r^2+5 \epsilon_0 r_s^2+6 \epsilon_s r^2\right)}{r_s \sqrt{10 \pi  \epsilon_0^2 r_s^2-9 \epsilon_0+9 \epsilon_s}}\right)\nonumber\\&& -\frac{2}{3} \log \left(8 \pi  \epsilon_0 r^2 \left(3 r^2-5 r_s^2\right)-24 \pi  \epsilon_s r^4+15 r_s^2\right)+\mathcal{C}\Bigg], \label{eq2.13}\\
  && \hspace{-0.5cm}    e^{-\mathcal{N}_0(r)}= \mathcal{B}_0(r)+\beta\, \mathcal{G} (r)=  \bigg[\frac{8 \pi  (\beta +1) r^2 \left\{3 r^2 (\epsilon_0-\epsilon_s)-5 \epsilon_0 r_s^2\right\}+15 r_s^2}{15 r_s^2}\bigg]. \label{eq2.14} 
\end{eqnarray}

Now the expressions for the components of new source $\theta_{ij}$ are written as: 
\begin{eqnarray}
  && \hspace{-0.2cm}  \theta^0_0 (r)= \frac{\epsilon_0 \left(r_s^2-r^2\right)+\epsilon_s r^2}{r_s^2}, \label{eq2.15}\\
  && \hspace{-0.2cm}  \theta^1_1 (r)= -\frac{\left(-3 \epsilon_0 r^2+5 \epsilon_0 r_s^2+3 \epsilon_s r^2\right) \left(r_s^2 \left(3-32 \pi   \mathcal{B}_g r^2\right)+8 \pi  \epsilon_0 r^2 \left(r_s^2-r^2\right)+8 \pi  \epsilon_s r^4\right)}{3 r_s^2 \left(-24 \pi  \epsilon_0 r^4+40 \pi  \epsilon_0 r^2 r_s^2+24 \pi  \epsilon_s r^4-15 r_s^2\right)},~~~~~~\label{eq2.16}\\
   && \hspace{-0.2cm}  \theta^2_2 (r)= \frac{1}{3 r_s^2 \left(8 \pi  \epsilon_0 r^2 \left(3 r^2-5 r_s^2\right)-24 \pi  \epsilon_s r^4+15 r_s^2\right)^2}\Big[\epsilon_0 \Big(5 r_s^4 \big[-768 \pi ^2  \mathcal{B}_g^2 r^6+64 \pi   \mathcal{B}_g r^4 \nonumber\\&&\hspace{0.8cm} \times \left(20 \pi   \mathcal{B}_g r_s^2+9\right)-18 r^2 \left(40 \pi   \mathcal{B}_g r_s^2+3\right)+45 r_s^2\big]+16 \pi  \epsilon_s r^4 r_s^2 \big(384 \pi   \mathcal{B}_g r^4-4 r^2\nonumber\\&&\hspace{0.8cm} \times \left(110 \pi   \mathcal{B}_g r_s^2+27\right)+105 r_s^2\big)-64 \pi ^2 \epsilon_s^2 r^8 \left(18 r^2-25 r_s^2\right)\Big)+6 \epsilon_s r^2 \big[5 r_s^4 \big(128 \pi ^2  \mathcal{B}_g^2 r^4\nonumber\\&&\hspace{0.8cm} -96 \pi   \mathcal{B}_g r^2+9\big)-16 \pi  \epsilon_s r^4 r_s^2 \left(32 \pi   \mathcal{B}_g r^2-9\right)+64 \pi ^2 \epsilon_s^2 r^8\big]+8 \pi  \epsilon_0^2 r^2 \big\{r_s^2 \big(-384 \pi   \mathcal{B}_g r^6\nonumber\\&&\hspace{0.8cm} +4 r^4 \left(220 \pi   \mathcal{B}_g r_s^2+27\right)-10 r^2 r_s^2 \left(40 \pi   \mathcal{B}_g r_s^2+21\right)+75 r_s^4\big)+16 \pi  \epsilon_s r^4 (9 r^4-25 r^2 r_s^2\nonumber\\&&\hspace{0.8cm}+20 r_s^4)\big\}-64 \pi ^2 \epsilon_0^3 r^4 \left(6 r^6-25 r^4 r_s^2+40 r^2 r_s^4-25 r_s^6\right)\Big]. \label{eq2.17}
\end{eqnarray}

\subsection{Mimicking of the Pressure Constraint: $P_r(r)=\theta^1_1(r)$} \label{solB}

By mimicking of the seed pressure ($P_r$) to component ($\theta^1_1$) through the Eqs. (\ref{eq20}) and (\ref{eq37}), we get directly exact solution for $\mathcal{G}(r)$ as
\begin{eqnarray}
    \mathcal{G}(r)=\frac{8 \pi  r^2 \left(-24 \pi  r_0 r^4+40 \pi  r_0 r^2 r_s^2+24 \pi  r_s r^4-15 r_s^2\right) \left\{-4 \mathcal{B}_g r_s^2+r_0 \left(r_s^2-r^2\right)+r_s r^2\right\}}{15 r_s^2 \left(r_s^2 \left(3-32 \pi  \mathcal{B}_g r^2\right)+8 \pi  r_0 r^2 \left(r_s^2-r^2\right)+8 \pi  r_s r^4\right)}.~~~~~~\label{eq2.18}
\end{eqnarray}

For this pressure constraint case, the deformed MGD solution can be described by the following spacetime geometries
\begin{eqnarray}
 && \hspace{-0.2cm}  e^{\mathcal{F}_0(r)}= e^{\mathcal{A}_0(r)}= \exp\Bigg[\frac{2 \sqrt{10 \pi } r_s (\epsilon_0-6  \mathcal{B}_g) }{3 \sqrt{10 \pi  \epsilon_0^2 r_s^2-9 \epsilon_0+9 \epsilon_s}}\tanh ^{-1}\left(\frac{\sqrt{\frac{2 \pi }{5}} \left(-6 \epsilon_0 r^2+5 \epsilon_0 r_s^2+6 \epsilon_s r^2\right)}{r_s \sqrt{10 \pi  \epsilon_0^2 r_s^2-9 \epsilon_0+9 \epsilon_s}}\right)\nonumber\\
 && \hspace{2.5cm}-\frac{2}{3} \log \left(8 \pi  \epsilon_0 r^2 \left(3 r^2-5 r_s^2\right)-24 \pi  \epsilon_s r^4+15 r_s^2\right)+\mathcal{C}\Bigg],\label{eq2.19}\\
  && \hspace{-0.2cm}    e^{-\mathcal{N}_0(r)}=   \beta \bigg[\frac{8 \pi  r^2 \left\{-24 \pi  r_0 r^4+40 \pi  r_0 r^2 r_s^2+24 \pi  r_s r^4-15 r_s^2\right\} \left\{-4 \mathcal{B}_g r_s^2+r_0 \left(r_s^2-r^2\right)+r_s r^2\right\}}{15 r_s^2 \left(r_s^2 \left(3-32 \pi  \mathcal{B}_g r^2\right)+8 \pi  r_0 r^2 \left(r_s^2-r^2\right)+8 \pi  r_s r^4\right)}\bigg]\nonumber\\
  &&\hspace{1.5cm}+ \bigg[\frac{24 \pi  r^4 (\epsilon_0-\epsilon_s)-40 \pi  \epsilon_0 r^2 r_s^2+15 r_s^2}{15 r_s^2}\bigg].  \label{eq2.20}
\end{eqnarray}

The expressions for the components of new source $\theta_{ij}$ as: 
\begin{eqnarray}
 &&\hspace{-0.3cm}   \theta^0_0(r)=\frac{1}{5 r_s^2 \left(r_s^2 \left(3-32 \pi  \mathcal{B}_g\,r^2\right)+8 \pi  \epsilon_0 r^2 \left(r_s^2-r^2\right)+8 \pi  \epsilon_s r^4\right)^2}\Big[\epsilon_0 \Big(r_s^4 \big[5120 \pi ^2 \mathcal{B}_g^2 r^6-32 \pi  \mathcal{B}_g\,r^4 \nonumber\\&&\hspace{0.8cm} \times \left(160 \pi  \mathcal{B}_g\,r_s^2+11\right)+15 r^2 \left(32 \pi  \mathcal{B}_g\,r_s^2-5\right)+45 r_s^2\big]-16 \pi  \epsilon_s r^4 r_s^2 \Big[320 \pi  \mathcal{B}_g\,r^4-2 r^2 \nonumber\\&&\hspace{0.8cm} \times  \left(160 \pi  \mathcal{B}_g\,r_s^2+11\right)+23 r_s^2\Big]+960 \pi ^2 \epsilon_s^2 r^8 \left(r^2-r_s^2\right)\Big)+\epsilon_s r^2 r_s^4 \big(-5120 \pi ^2 \mathcal{B}_g^2 r^4+352 \pi  \mathcal{B}_g\,r^2\nonumber\\&&\hspace{0.8cm} +75\big)-16 \pi  \epsilon_0^2 r^2 \Big[r_s^2 \left(-160 \pi  \mathcal{B}_g\,r^6+r^4 \left(320 \pi  \mathcal{B}_g\,r_s^2+11\right)-r^2 r_s^2 \left(160 \pi  \mathcal{B}_g\,r_s^2+23\right)+10 r_s^4\right)\nonumber\\&&\hspace{0.8cm} +60 \pi  \epsilon_s r^4 \left(r^2-r_s^2\right)^2\Big]+16 \pi  \epsilon_s^2 r^6 r_s^2 \left(160 \pi  \mathcal{B}_g\,r^2-11\right)+20 \mathcal{B}_g\,r_s^6 \left(32 \pi  \mathcal{B}_g\,r^2-9\right)+320 \pi ^2 \epsilon_0^3 r^4 \nonumber\\&&\hspace{0.8cm} \times \left(r^2-r_s^2\right)^3-320 \pi ^2 \epsilon_s^3 r^{10}\Big], \label{eq2.21}\\
  &&\hspace{-0.3cm}   \theta^1_1(r)=\frac{-4 \mathcal{B}_g\,r_s^2+\epsilon_0 \left(r_s^2-r^2\right)+\epsilon_s r^2}{3 r_s^2},\label{eq2.22}\\
    &&\hspace{-0.3cm} \theta^2_2(r)= \Big[6 \pi  \mathcal{G}_{11} (r) r^2 \left(8 \pi  \epsilon_0 r^2 \left(3 r^2-5 r_s^2\right)-24 \pi  \epsilon_s r^4+15 r_s^2\right) \left(4 \mathcal{B}_g\,r_s^2+\epsilon_0 \left(r^2-r_s^2\right)-\epsilon_s r^2\right)\nonumber\\&&\hspace{0.8cm}+\mathcal{G}_{22} (r) \left(-24 \pi  \epsilon_0 r^4+40 \pi  \epsilon_0 r^2 r_s^2+24 \pi  \epsilon_s r^4-15 r_s^2\right) \Big(5 r_s^2 \left(16 \pi  \mathcal{B}_g\,r^2-3\right)+8 \pi  \epsilon_0 r^4\nonumber\\&&\hspace{0.8cm}-8 \pi  \epsilon_s r^4\Big)\Big] \Big[15 r_s^2 \left(8 \pi  \epsilon_0 r^2 \left(3 r^2-5 r_s^2\right)-24 \pi  \epsilon_s r^4+15 r_s^2\right)^2 \Big(r_s^2 \left(3-32 \pi  \mathcal{B}_g\,r^2\right)+8 \pi  \epsilon_0 r^2 \nonumber\\&&\hspace{0.8cm} \times \left(r_s^2-r^2\right)+8 \pi  \epsilon_s r^4\Big)^2\Big]^{-1},~~~~~~~~~~\label{eq2.2.23}
\end{eqnarray}
where the expressions for the coefficients used in the above expressions are: 
\begin{eqnarray}
   &&\hspace{-0.4cm} \mathcal{G}_{11} (r)=\left[r_s^2 \left(32 \pi  \mathcal{B}_g\,r^2-3\right)+8 \pi  \epsilon_0 r^2 \left(r^2-r_s^2\right)-8 \pi  \epsilon_s r^4\right] \big[5 \epsilon_0 r_s^2 \big(112 \pi  \mathcal{B}_g\,r^4-8 r^2 (10 \pi  \mathcal{B}_g\,r_s^2\nonumber\\&&\hspace{0.8cm}+3)+15 r_s^2\big)+40 \epsilon_s r^2 r_s^2 \left(3-14 \pi  \mathcal{B}_g\,r^2\right)+50 \mathcal{B}_g\,r_s^4 \left(8 \pi  \mathcal{B}_g\,r^2-3\right)+4 \pi  \epsilon_0^2 r^2 (16 r^4-30 r^2 r_s^2\nonumber\\&&\hspace{0.8cm} +25 r_s^4)-8 \pi  \epsilon_0 \epsilon_s r^4 \left(16 r^2-15 r_s^2\right)+64 \pi  \epsilon_s^2 r^6\big],\nonumber\\
  &&\hspace{-0.5cm}\mathcal{G}_{22} (r)=\epsilon_0 \Big[r_s^4 \left(6144 \pi ^2 \mathcal{B}_g^2 r^6-32 \pi  \mathcal{B}_g\,r^4 \left(160 \pi  \mathcal{B}_g\,r_s^2+27\right)+30 r^2 \left(32 \pi  \mathcal{B}_g\,r_s^2-3\right)+45 r_s^2\right)\nonumber\\&&\hspace{0.8cm} -64 \pi  \epsilon_s r^4 r_s^2 \left(96 \pi  \mathcal{B}_g\,r^4-r^2 \left(88 \pi  \mathcal{B}_g\,r_s^2+9\right)+9 r_s^2\right)+64 \pi ^2 \epsilon_s^2 r^8 \left(18 r^2-17 r_s^2\right)\Big]\nonumber\\&&\hspace{0.8cm} -6 \Big[\epsilon_s r^2 r_s^4 \left(1024 \pi ^2 \mathcal{B}_g^2 r^4-144 \pi  \mathcal{B}_g\,r^2-15\right)+16 \pi  \epsilon_s^2 r^6 r_s^2 \left(3-32 \pi  \mathcal{B}_g\,r^2\right)+30 \mathcal{B}_g\,r_s^6\nonumber\\&&\hspace{0.8cm}+64 \pi ^2 \epsilon_s^3 r^{10}\Big]-16 \pi  \epsilon_0^2 r^2 \Big[r_s^2 \big(-192 \pi  \mathcal{B}_g\,r^6+2 r^4 \left(176 \pi  \mathcal{B}_g\,r_s^2+9\right)-4 r^2 r_s^2 \left(40 \pi  \mathcal{B}_g\,r_s^2+9\right)\nonumber\\&&\hspace{0.8cm} +15 r_s^4\big)+8 \pi  \epsilon_s r^4 \left(9 r^4-17 r^2 r_s^2+8 r_s^4\right)\Big]+64 \pi ^2 \epsilon_0^3 r^4 \left(r^2-r_s^2\right)^2 \left(6 r^2-5 r_s^2\right). \nonumber
\end{eqnarray}

\section{Boundary conditions for SS models under Gravitational decoupling}\label{sec4} 

To establish a realistic compact stellar model, characterized by a confined and bounded distribution of matter with a well defined mass $M$ and radius $R$, it is essential to connect the inner geometry $\mathcal{M^{-}}$ at the surface $\Sigma=r=R$ with the outside space-time $\mathcal{M^{+}}$ that surrounds the structure. In the context of the GR system, the exterior manifolds is widely recognized as that of Schwarzschild vacuum space-time, specifically when considering uncharged, non-radiating, and static compact objects. Nevertheless, it is necessary to examine the characteristics of the additional component of the energy-momentum tensor, namely the $\theta$-sector. Furthermore, this novel additional term has the potential to alter the material composition of the exterior space-time. The precise geometry that characterizes this exterior manifold can be provided as
\begin{equation}
ds^{2}=\left[1-\frac{2{\mathcal{M}}}{r}\right]dt^{2}- \frac{dr^2}{ 1-\frac{2{\mathcal{M}}}{r}+\beta\, \mathcal{G^\ast}(r)} -r^{2}d\Omega^{2}. \label{eq3.1}
\end{equation}

The geometric deformation function for the outside Schwarzschild space-time caused by the $\theta_{ij}$ source is called $\mathcal{G^\ast}(r)$. This metric (\ref{eq3.1}) denotes a distorted Schwarzschild space-time that is no longer a vacuum. Nevertheless, it is possible to make $\mathcal{G^\ast}(r)$ null in order to preserve the typical outer vacuum space-time, without compromising its generality and benefiting from simplicity.  To integrate the internal configuration with the external one effectively, the ID junction conditions need the use of both the first and second basic forms. The first basic form determines the continuity of the metric potentials at the $\Sigma$ boundary. The first basic form is as follows:
\begin{equation}\label{eq3.2}
e^{\mathcal{A}_0^{-}(r)}|_{r=r_s}=e^{\mathcal{A}_0^{+}(r)}|_{r=r_s},
\end{equation}
and
\begin{equation}\label{eq3.3}
e^{\mathcal{F}_0^{-}(r)}|_{r=r_s}=e^{\mathcal{F}_0^{+}(r)}|_{r=r_s},
\end{equation}
where the symbols $``-"$ and $``+"$ represent the inner and outer geometries, respectively. The second basic form pertains to the continuity of the extrinsic curvature $K_{\mu\nu}$ caused by the components $\mathcal{M}^{-}$ and $\mathcal{M}^{+}$ on the  surface. The continuity of the $K_{rr}$ component throughout the field $\Sigma$ results in
\begin{equation}\label{eq3.4}
\left[P_r^{(\text{eff})}(r)\right]_{\Sigma}=\left[P^{(\text{eff})}_r(r)-\beta\, \theta^{1}_{r}(1)\right]_{\Sigma}=0.
\end{equation} 
which yields
  \begin{equation}
{P_r}(r_s)+\beta\,(\theta^1_1)^{-}(r_s)=\beta\,(\theta^1_1)^{+}(r_s).~~~\label{eq3.5}
  \end{equation}
  
Referring to Eqs. (\ref{eq37}) and (\ref{eq3.5}), we may now deduce the following result:
\begin{eqnarray}
{P}_r(r_s)+\beta\,\bigg[\mathcal{G}(r_s)\left(\frac{\mathcal{F}_0^{\prime}(r_s)}{r_s}+\frac{1}{r_s^{2}}\right)\bigg] =\beta\,(\theta^1_1)^{+}(r_s).~~\label{eq3.6}
\end{eqnarray}

The substitution of the outside space-time in Eq. (\ref{eq37}) with Eq. (\ref{eq3.6}) yields
\begin{eqnarray}
 P_r(r_s)+\beta\,\bigg[\mathcal{G}(r_s)\left(\frac{\mathcal{F}_0^{\prime}(r_s)}{r_s}+\frac{1}{r_s^{2}}\right)\bigg] =\beta\,\mathcal{G}^{\ast}(r_s)\Bigg[\frac{2\mathcal{M}}{r^3_s\,\Big(1-\frac{2\mathcal{M}}{r_s}\Big)}+\frac{1}{r_s^2}\Bigg].~~~\label{eq3.7}
 \end{eqnarray} 
 
The equations (\ref{eq3.2}), (\ref{eq3.3}) and (\ref{eq3.7}) are necessary conditions for the establishment of a spherically symmetric external "vacuum" specified by the deformed Schwarzschild de-Sitter metric in Eq. (\ref{eq3.1}) in relation to the interior MGD metric (\ref{eq4}). The external structure of the system could be comprised of fields that are defined in the source $\theta_{ij}$. A fundamental second form is derived from the matching condition (\ref{eq3.7}): if the outer geometry is defined by the exact Schwarzschild metric, then it is necessary to have $\mathcal{G}^{\ast}(r)=0$ in Eq.~(\ref{eq3.1}), which therefore results in the condition
\begin{eqnarray}
 P^{\text{eff}}_r(r_s)=P_r(r_s)+\beta\,\bigg[\mathcal{G}(r_s)\left(\frac{\mathcal{F}_0^{\prime}(r_s)}{r_s}+\frac{1}{r_s^{2}}\right)\bigg] =0.~~ \label{eq3.8} 
 \end{eqnarray}  
 
 Now we shall use the conditions (\ref{eq3.2}), (\ref{eq3.3}) and (\ref{eq3.8}) to derive the arbitrary constant involved in the solution.  Through these conditions, the expression for constants corresponding to solutions (\ref{solA}) and (\ref{solB}) are given in the below subsections. 

\subsection{The expressions of the constant for the solution~\ref{solA}:}
\begin{eqnarray}
&&\hspace{-0.5cm}    \mathcal{B}_g=\frac{6 \beta  r_0+r_s \left\{9 \beta +16 \pi  (\beta +1) r_0 r_s^2-15\right\}+24 \pi  (\beta +1) r_s^2 r_s^2}{64 \pi  (\beta +1) r_0 r_s^2+96 \pi  (\beta +1) r_s r_s^2-60},\\
&&\hspace{-0.5cm}    \mathcal{C}= \frac{2 \sqrt{10 \pi } \,r_s (\epsilon_0-6 \mathcal{B}_g) }{3 \sqrt{10 \pi  \epsilon_0^2 r_s^2-9 \epsilon_0+9 \epsilon_s}} \tanh ^{-1}\left(\frac{\sqrt{\frac{2 \pi }{5}} \,r_s (\epsilon_0-6 \epsilon_s)}{\sqrt{10 \pi  \epsilon_0^2 r_s^2-9 \epsilon_0+9 \epsilon_s}}\right)+\frac{2}{3} \ln \left[15 r_s^2-8 \pi  r_s^4 (2 \epsilon_0+3 \epsilon_s)\right] \nonumber\\
&&\hspace{0.8cm}+\ln \bigg[1-\frac{1}{15} 16 \pi  (\beta +1) \epsilon_0 r_s^2-\frac{8}{5} \pi  (\beta +1) \epsilon_s r_s^2\bigg].
\end{eqnarray}

\subsection{The expressions of the constant for the solution~\ref{solB}:}
\begin{eqnarray}
&&\hspace{-0.5cm}    \mathcal{B}_g=\frac{\epsilon_s}{32 \pi },\\
&&\hspace{-0.5cm}    \mathcal{C}= \frac{2 \sqrt{10 \pi } \,r_s (\epsilon_0-6 \mathcal{B}_g) }{3 \sqrt{10 \pi  \epsilon_0^2 r_s^2-9 \epsilon_0+9 \epsilon_s}} \tanh ^{-1}\left(\frac{\sqrt{\frac{2 \pi }{5}} \,r_s (\epsilon_0-6 \epsilon_s)}{\sqrt{10 \pi  \epsilon_0^2 r_s^2-9 \epsilon_0+9 \epsilon_s}}\right)+\frac{2}{3} \ln \left[15 r_s^2-8 \pi  r_s^4 (2 \epsilon_0+3 \epsilon_s)\right]\nonumber\\
&&\hspace{0.5cm}+\ln \left[\frac{\left(16 \pi  \epsilon_0 r_s^2+24 \pi  \epsilon_s r_s^2-15\right) \left\{32 \pi  (\beta -1) \mathcal{B}_g r_s^2-8 \pi  (\beta -1) \epsilon_s r_s^2+3\right\}}{15 \left(32 \pi  \mathcal{B}_g r_s^2-8 \pi  \epsilon_s r_s^2-3\right)}\right].
\end{eqnarray}

\begin{table*}[!htp]
\footnotesize
\centering
\caption{The predicted radii of few high mass CSs corresponding to Fig. \ref{fig5a}.}\label{tab1}
\scalebox{0.82}{\begin{tabular}{|*{10}{c|} }
\hline
& & \multicolumn{8}{c|}{{Predicted $R$ (km) in $\epsilon=\theta_0^0$}} \\[0.15cm]
\cline{3-10}
{Objects} & {$\frac{M}{M_\odot}$} & \multicolumn{5}{c|}{ $\beta$} & \multicolumn{3}{c|}{$\mathcal{B}_g~~(MeV/fm^3)$} \\[0.15cm]
\cline{3-7} \cline{8-10}
&  & 0 & $0.025$ & $0.050$ & $0.075$ & $0.10$ & $55$ & $60$ & $65$ \\[0.15cm] \hline
PSR J1614-2230 \citep{Demorest:2010bx}  &  $1.97 \pm 0.04$  & $10.86_{-0.05}^{+0.03}$  &  $10.91_{-0.05}^{+0.04}$  &   $10.96_{-0.06}^{+0.04}$  &  $10.99_{-0.06}^{+0.05}$  &  $11.03_{-0.05}^{+0.06}$ & $11.50_{-0.06}^{+0.05}$ & $11.28_{-0.04}^{+0.04}$  & $10.60_{-0.21}^{+0.09}$   \\[0.15cm]
\hline
PSR J0952-0607 \citep{Romani:2022jhd} & $2.35 \pm 0.17$  & $11.05_{-0.02}^{-}$  &  $11.20_{-0.09}^{+0.01}$  &   $11.30_{-0.11}^{+0.03}$  &  $11.39_{-0.15}^{+0.08}$  &  $11.46_{-0.17}^{+0.15}$ & $11.93_{-0.16}^{+0.11}$ & $11.40_{-}^{0.04}$  & $-$   \\ [0.15cm]
\hline
GW190814 \citep{LIGOScientific:2020zkf} & $2.5-2.67$ & $-$ &  $10.93_{-}^{+0.21}$  &  $11.31_{-0.25}^{+0.03}$  &  $11.47_{-0.01}^{+0.01}$ & $11.59_{-0.03}^{+0.02}$ & $12.07_{-0.04}^{+0.01}$  & $-$ & $-$   \\[0.15cm]
\hline
GW200210 \citep{KAGRA:2021vkt} & $2.83^{+0.47}_{-0.42}$ &  $-$  &  $-$  & $-$  & $-$ &  $11.45_{-}^{+0.05}$ & $11.83_{-}^{+0.15}$ & $-$  & $-$  \\[0.15cm]
\hline
\end{tabular} }
 \end{table*}

\begin{table*}[!htp]
\footnotesize
\centering

\caption{The predicted moment of inertia of few high mass CSs corresponding to Fig. \ref{fig5a}.}\label{tab2}
\scalebox{1.1}{\begin{tabular}{|*{10}{c|} }
\hline
& & \multicolumn{8}{c|}{{Predicted $I$ ($\times 10^{45}g\cdot cm^2$) in $\epsilon=\theta_0^0$}} \\[0.15cm]
\cline{3-10}
{Objects} & {$\frac{M}{M_\odot}$} & \multicolumn{5}{c|}{ $\beta$} & \multicolumn{3}{c|}{$\mathcal{B}_g~~(MeV/fm^3)$} \\[0.15cm]
\cline{3-7} \cline{8-10}
&  & 0 & $0.025$ & $0.050$ & $0.075$ & $0.10$ & $55$ & $60$ & $65$ \\[0.15cm] \hline
PSR J1614-2230 \citep{Demorest:2010bx}  &  $1.97 \pm 0.04$  & $1.87$  &  $1.88$  &   $1.89$  &  $1.90$  &  $1.91$ & $2.04$ & $1.98$  & $1.80$   \\[0.15cm]
\hline
PSR J0952-0607 \citep{Romani:2022jhd} & $2.35 \pm 0.17$  & $2.49$  &  $2.55$  &   $2.59$  &  $2.61$  &  $2.63$ & $2.80$ & $2.62$  & $-$   \\ [0.15cm]
\hline
GW190814 \citep{LIGOScientific:2020zkf} & $2.5-2.67$ & $-$ &  $-$  &  $2.98$  &  $3.05$ & $3.10$ & $3.29$  & $-$ & $-$   \\[0.15cm]
\hline
GW200210 \citep{KAGRA:2021vkt} & $2.83^{+0.47}_{-0.42}$ &  $-$  &  $-$  & $-$  & $-$ &  $3.50$ & $3.67$ & $-$  & $-$  \\[0.15cm]
\hline
\end{tabular} }
 \end{table*}

\begin{table*}[!htp]
\footnotesize
\centering
\caption{The predicted radii of few high mass CSs corresponding to Fig. \ref{fig5b}.}\label{tab3}
\scalebox{0.82}{\begin{tabular}{|*{10}{c|} }
\hline
& & \multicolumn{8}{c|}{{Predicted $R$ (km) in $P_r=\theta_1^1$}} \\[0.15cm]
\cline{3-10}
{Objects} & {$\frac{M}{M_\odot}$} & \multicolumn{5}{c|}{ $\beta$} & \multicolumn{3}{c|}{$\mathcal{B}_g~~(MeV/fm^3)$} \\[0.15cm]
\cline{3-7} \cline{8-10}
&  & 0 & $0.025$ & $0.050$ & $0.075$ & $0.10$ & $55$ & $60$ & $65$ \\[0.15cm] \hline
PSR J1614-2230 \citep{Demorest:2010bx}  &  $1.97 \pm 0.04$  & $11.06_{-0.06}^{+0.05}$  &  $11.04_{-0.06}^{+0.05}$  &   $11.02_{-0.05}^{+0.04}$  &  $11.00_{-0.06}^{+0.04}$  &  $10.98_{-0.06}^{+0.05}$ & $11.09_{-0.06}^{+0.06}$ & $11.06_{-0.06}^{+0.04}$  & $10.99_{-0.05}^{+0.05}$   \\[0.15cm]
\hline
PSR J0952-0607 \citep{Romani:2022jhd} & $2.35 \pm 0.17$  & $11.51_{-0.08}^{+0.13}$  &  $11.47_{-0.17}^{+0.12}$  &   $11.44_{-0.16}^{+0.10}$  &  $11.40_{-0.15}^{+0.09}$  &  $11.36_{-0.14}^{+0.06}$ & $11.55_{-0.18}^{+0.14}$ & $11.46_{-0.15}^{0.03}$  & $11.11_{-}^{0.08}$   \\ [0.15cm]
\hline
GW190814 \citep{LIGOScientific:2020zkf} & $2.5-2.67$ & $11.67_{-0.05}^{+0.04}$ &  $11.61_{-0.03}^{+0.02}$  &  $11.56_{-0.03}^{+0.01}$  &  $11.49_{-0.01}^{+0.05}$ & $11.41_{-0.15}^{+0.02}$ & $11.72_{-0.04}^{+0.02}$  & $11.42_{-}^{+0.08}$ & $-$   \\[0.15cm]
\hline
GW200210 \citep{KAGRA:2021vkt} & $2.83^{+0.47}_{-0.42}$ &  $11.68_{-0.12}^{-}$  &  $11.50_{-}^{+0.02}$  & $-$  & $-$ &  $-$ & $-$ & $-$  & $-$  \\[0.15cm]
\hline
\end{tabular} }
\end{table*}

\begin{table*}[!htp]
\footnotesize
\centering
\begin{minipage}{0.8\textheight}
\caption{The predicted moment of inertia of few high mass CSs corresponding to Fig. \ref{fig5b}.}\label{tab4}
\scalebox{1.1}{\begin{tabular}{|*{10}{c|} }
\hline
& & \multicolumn{8}{c|}{{Predicted $I$ ($\times 10^{45} g\cdot cm^2$) in $P_r=\theta_1^1$}} \\[0.15cm]
\cline{3-10}
{Objects} & {$\frac{M}{M_\odot}$} & \multicolumn{5}{c|}{ $\beta$} & \multicolumn{3}{c|}{$\mathcal{B}_g~~(MeV/fm^3)$} \\[0.15cm]
\cline{3-7} \cline{8-10}
&  & 0 & $0.025$ & $0.050$ & $0.075$ & $0.10$ & $55$ & $60$ & $65$ \\[0.15cm] \hline
PSR J1614-2230 \citep{Demorest:2010bx}  &  $1.97 \pm 0.04$  & $1.93$  &  $1.92$  &   $1.91$  &  $1.90$  &  $1.89$ & $1.94$ & $1.93$  & $1.91$   \\[0.15cm]
\hline
PSR J0952-0607 \citep{Romani:2022jhd} & $2.35 \pm 0.17$  & $2.66$  &  $2.64$  &   $2.63$  &  $2.61$  &  $2.60$ & $2.67$ & $2.63$  & $2.52$   \\ [0.15cm]
\hline
GW190814 \citep{LIGOScientific:2020zkf} & $2.5-2.67$ & $3.13$ &  $3.11$  &  $3.08$  &  $3.05$ & $3.02$ & $3.15$  & $3.00$ & $-$   \\[0.15cm]
\hline
GW200210 \citep{KAGRA:2021vkt} & $2.83^{+0.47}_{-0.42}$ &  $3.59$  &  $3.52$  & $-$  & $-$ &  $-$ & $-$ & $-$  & $-$  \\[0.15cm]
\hline
\end{tabular} }
 \end{minipage}
\end{table*}

\begin{table*}[!htp]
\footnotesize
\centering
\caption{Maximum masses for the case $\epsilon = \theta_0^0$ and $P_r = \theta_1^1$ from the $M-R$ curves (in Figs. \ref{fig5a} and \ref{fig5b}).}\label{tab5}
\scalebox{1.2}{\begin{tabular}{|*{10}{c|}}
\hline
\multicolumn{10}{|c|}{{$M_{max}/M_\odot$ and $R$ (km) in $\epsilon=\theta_0^0$}} \\[0.15cm]
\hline
& \multicolumn{5}{|c|}{ $\beta$} & \multicolumn{4}{c|}{$\mathcal{B}_g~~(MeV/fm^3)$}\\[0.15cm]
\cline{1-7} \cline{8-10}
& 0 & $0.025$ & $0.050$ & $0.075$ & $0.10$ & 55 & 60 & 65 & 70 \\[0.15cm] \hline
$M_{max}$ & 2.48  &  2.58  &  2.67 &  2.77 & 2.87 & 2.85 & 2.43 & 2.01 & 1.58   \\
$R$ & 10.69  &  10.83  &  10.92 &  11.07 & 11.20 & 11.68 & 11.07 & 10.36 & 9.64   \\[0.15cm]
\hline
\multicolumn{10}{|c|}{{$M_{max}/M_\odot$ and $R$ (km) in $P_r=\theta_1^1$}} \\[0.15cm]
\hline
& \multicolumn{5}{|c|}{ $\beta$} & \multicolumn{4}{c|}{$\mathcal{B}_g~~(MeV/fm^3)$}\\[0.15cm]
\cline{1-7} \cline{8-10}
& 0 & $0.025$ & $0.050$ & $0.075$ & $0.10$ & 55 & 60 & 65 & 70 \\[0.15cm] \hline
$M_{max}$ & 2.95  &  2.87  &  2.80 &  2.74 & 2.69 & 2.81 & 2.59 & 2.37 & 2.14   \\[0.15cm] 
$R$ & 11.32 & 11.29  &  11.20  &  11.14 &  11.09 & 11.46 & 11.26 & 10.94 & 10.67    \\[0.15cm]
\hline
\end{tabular} }
 \end{table*}

\begin{figure}[!htp]
    \centering
\includegraphics[height=5.8cm,width=6.0cm]{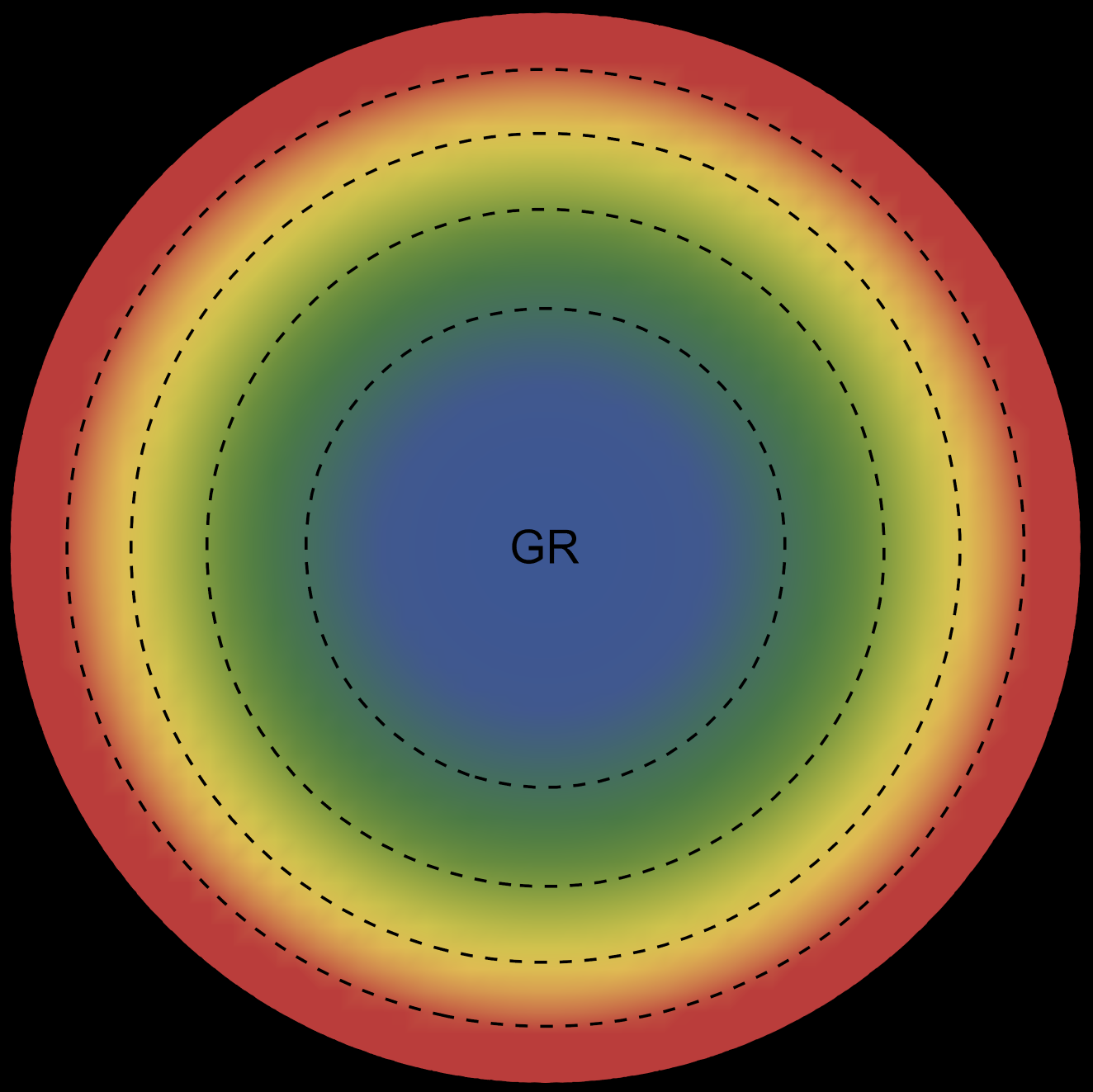}~~\includegraphics[height=5.5cm,width=1.5cm]{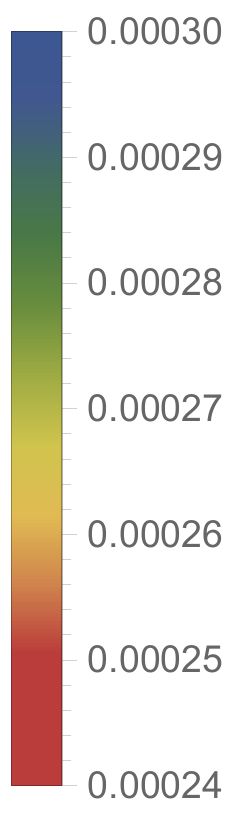}~~~~~~\includegraphics[height=5.8cm,width=6.0cm]{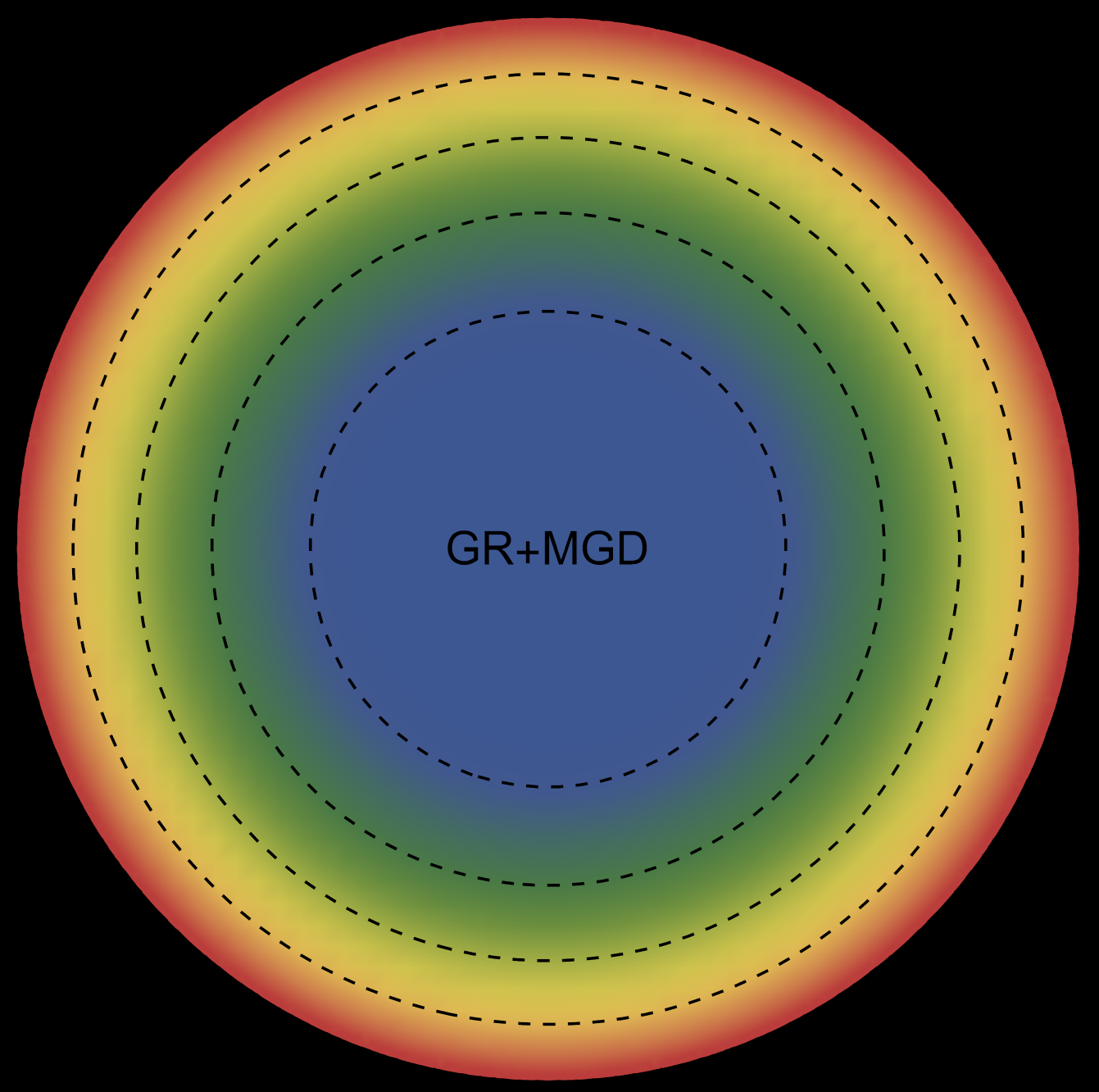}~~\includegraphics[height=5.5cm,width=1.5cm]{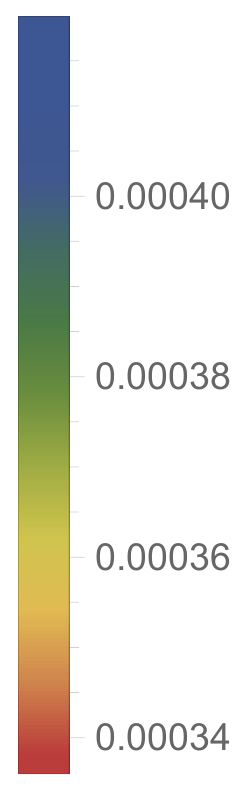}
    \caption{Density ($\rho^{\text{eff}} (r)$) profile within the stellar object corresponding to GR ($\beta=0$) and GR+MGD ($\beta=0.1$) for the parameter values $\epsilon_0=0.0003~\text{km}^{-2}$, $\epsilon_s= 0.00024~\text{km}^{-2}$ and $r_s=11.5~\text{km}$ for solution~\ref{solA} $(\epsilon=\theta^0_0)$.}
    \label{fig1a}
\end{figure}

\begin{figure}[!htp]
    \centering
\includegraphics[height=5.8cm,width=6.0cm]{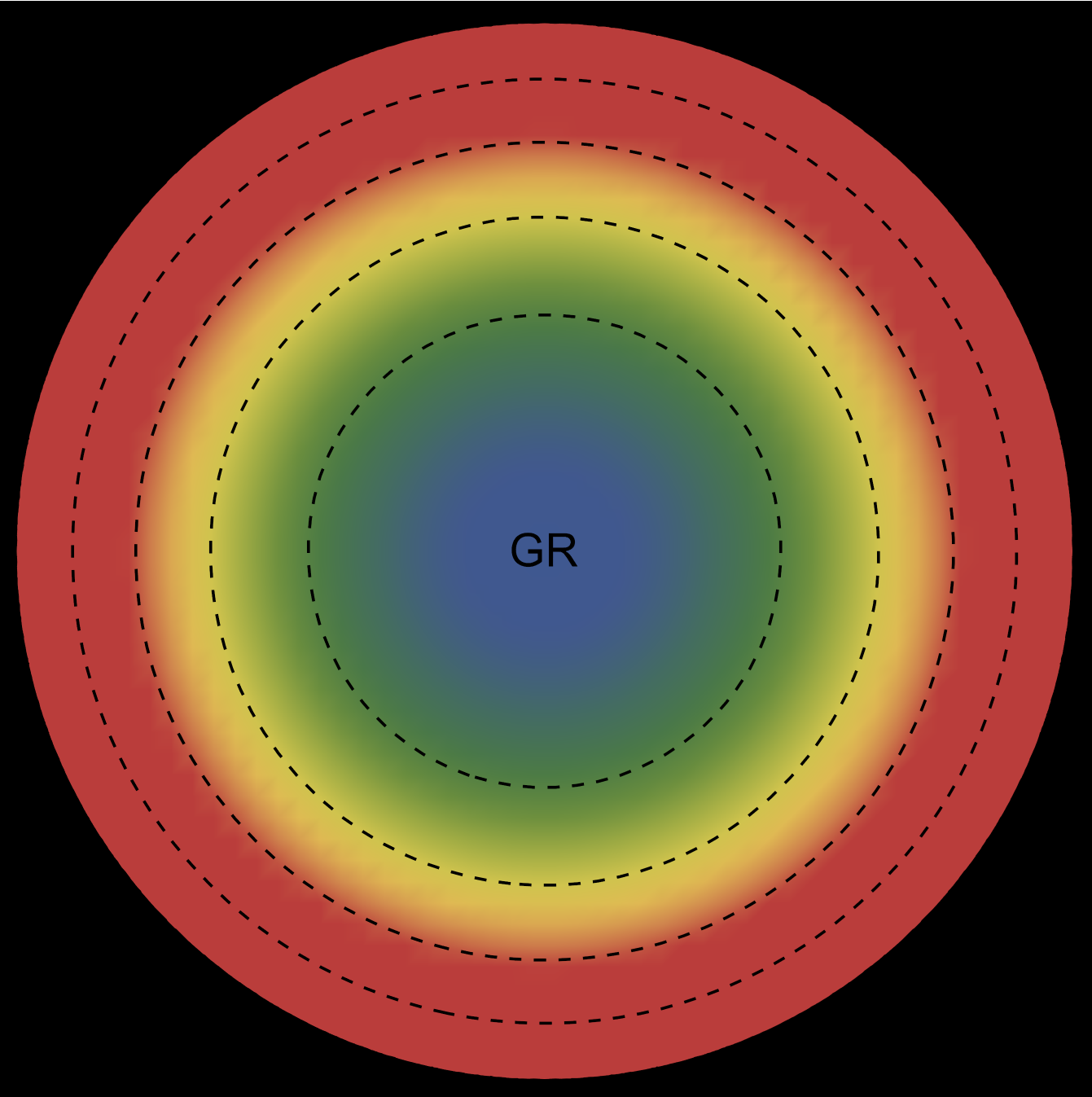}~~\includegraphics[height=5.5cm,width=1.5cm]{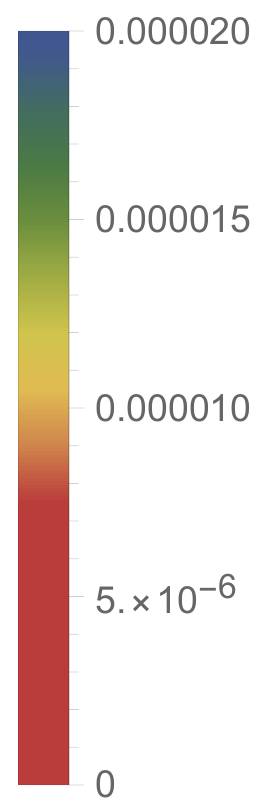}~~~~~~\includegraphics[height=5.8cm,width=6.0cm]{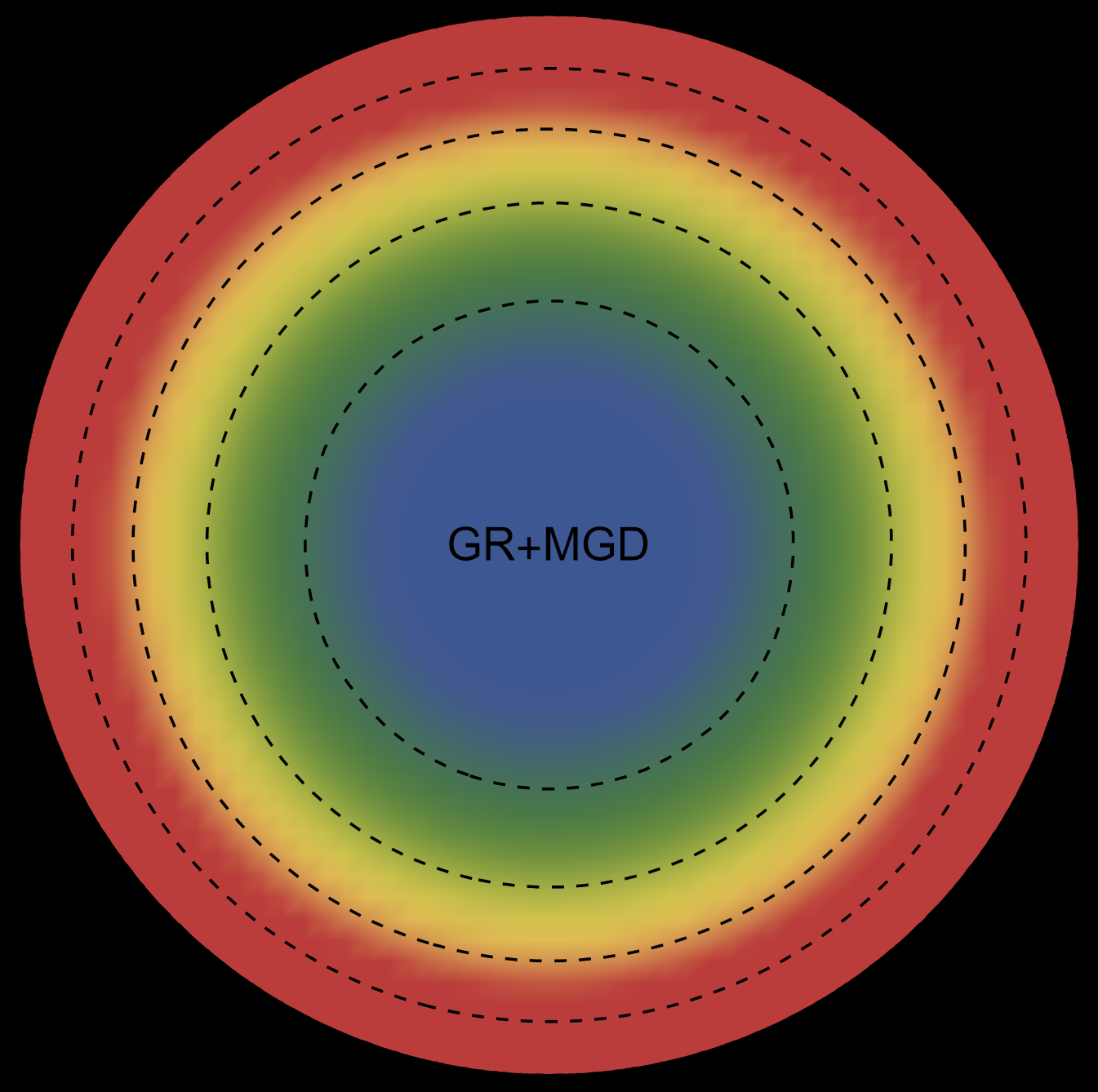}~~\includegraphics[height=5.5cm,width=1.5cm]{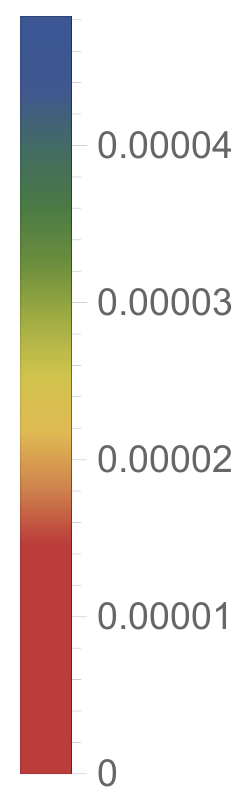}
    \caption{Radial pressure ($P^{\text{eff}}_r (r)$) profile  within the stellar object corresponding to GR ($\beta=0$) and GR+MGD ($\beta=0.1$) for the parameter values $\epsilon_0=0.0003~\text{km}^{-2}$, $\epsilon_s= 0.00024~\text{km}^{-2}$ and $r_s=11.5~km$}
    \label{fig2a}
\end{figure}

\begin{figure}[!htp]
    \centering
\includegraphics[height=5.8cm,width=6.0cm]{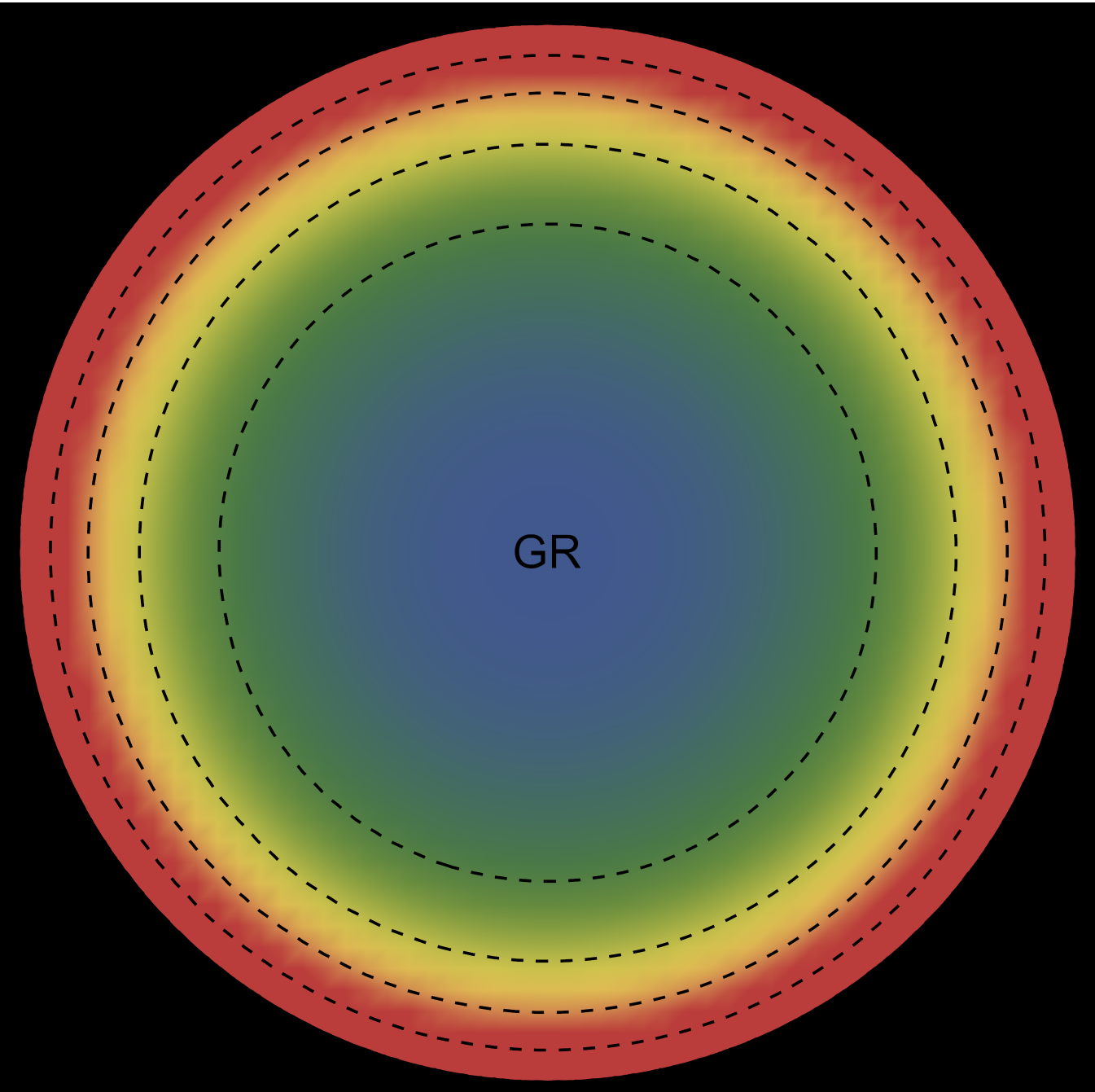}~~\includegraphics[height=5.5cm,width=1.5cm]{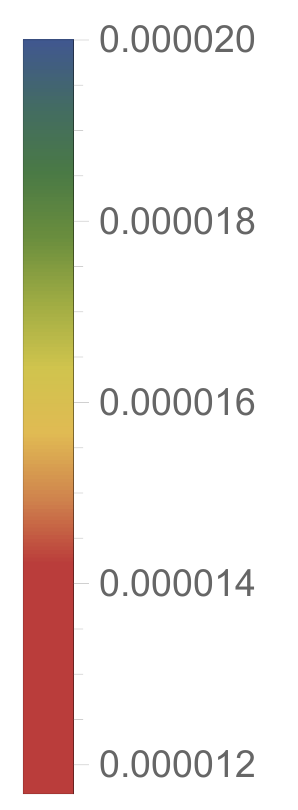}~~~~~~\includegraphics[height=5.8cm,width=6.0cm]{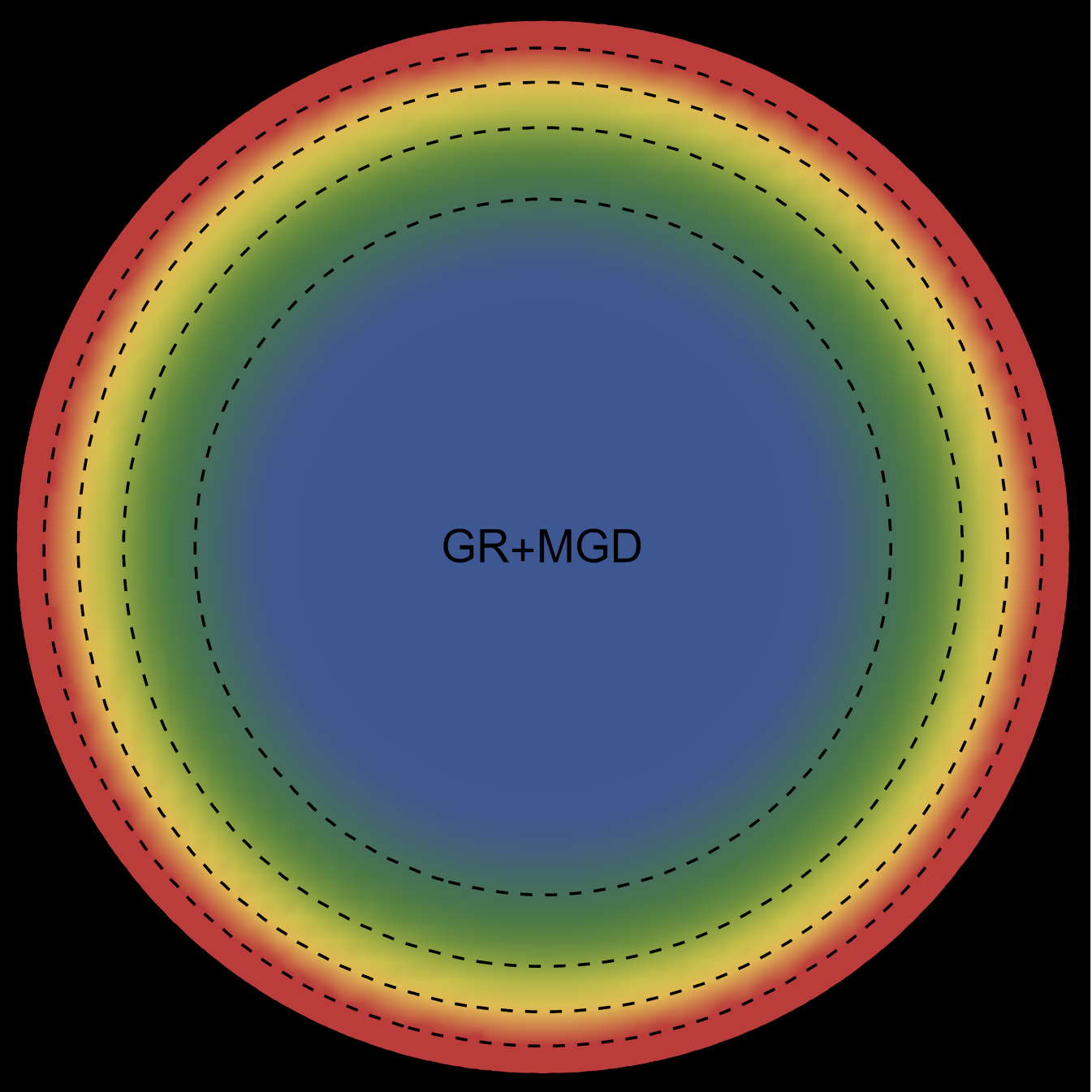}~~\includegraphics[height=5.5cm,width=1.5cm]{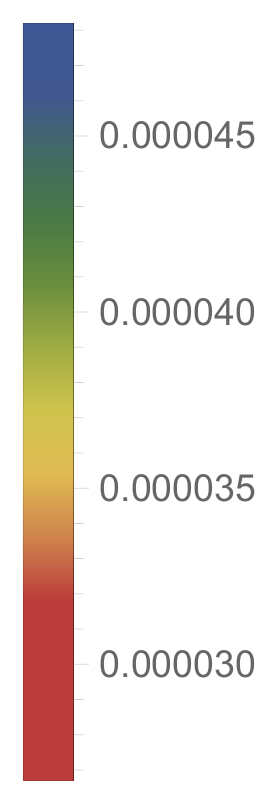}
    \caption{Tangential pressure ($P^{\text{eff}}_t (r)$) profile  within the stellar object corresponding to GR ($\beta=0$) and GR+MGD ($\beta=0.1$) for the parameter values $\epsilon_0=0.0003~\text{km}^{-2}$, $\epsilon_s= 0.00024~\text{km}^{-2}$ and $r_s=11.5~\text{km}$ for solution~\ref{solA} $(\epsilon=\theta^0_0)$.}
    \label{fig3a}
\end{figure}

\begin{figure}[!htp]
    \centering
\includegraphics[height=5.8cm,width=6.0cm]{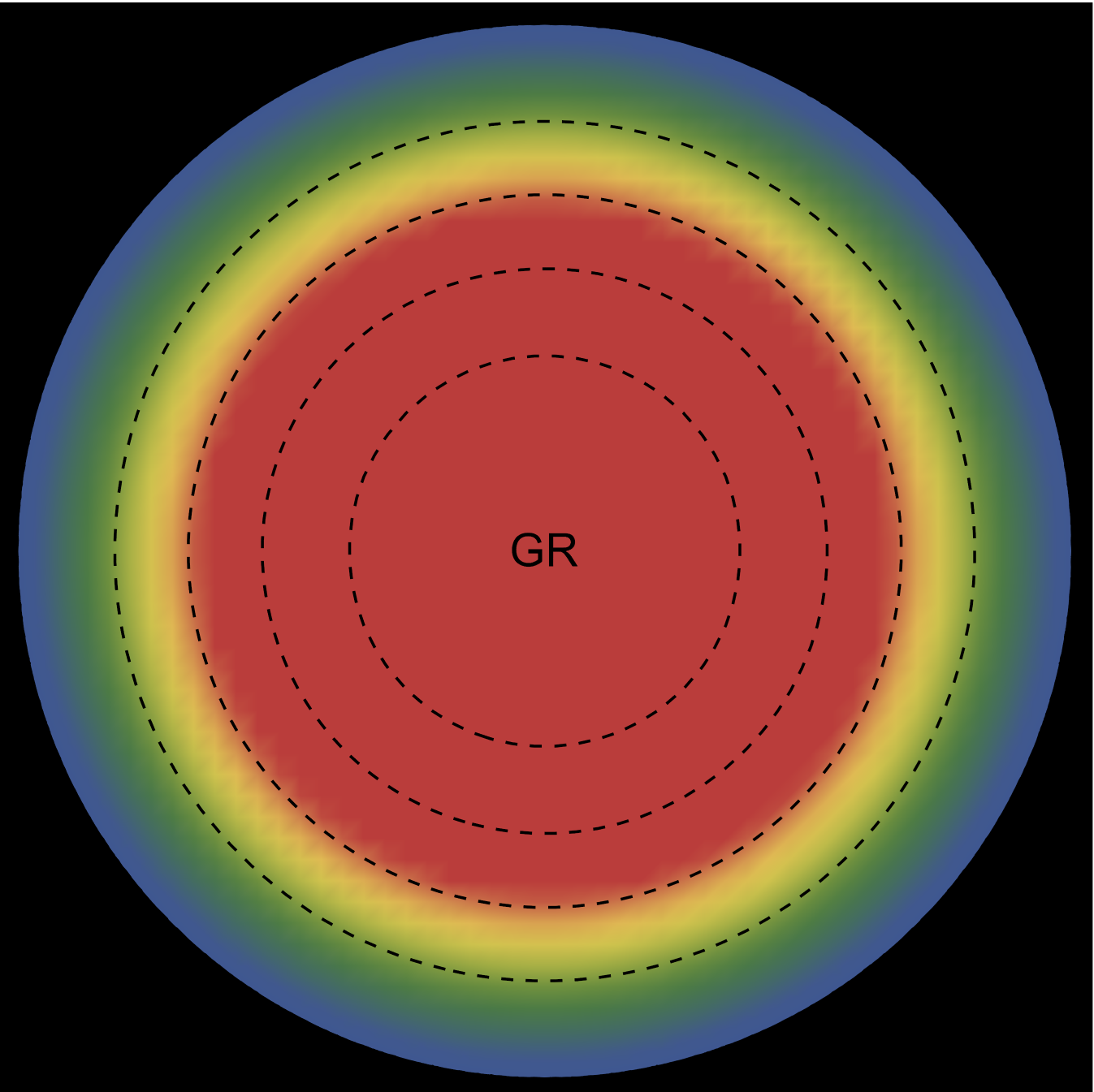}~~\includegraphics[height=5.5cm,width=1.5cm]{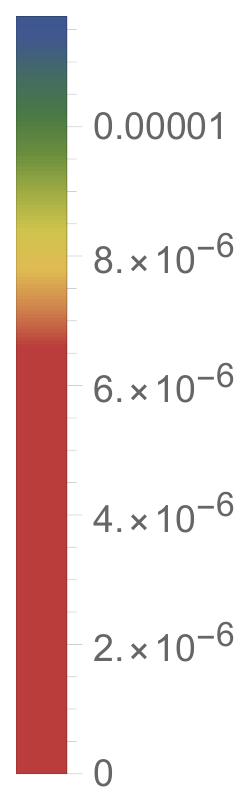}~~~~~~\includegraphics[height=5.8cm,width=6.0cm]{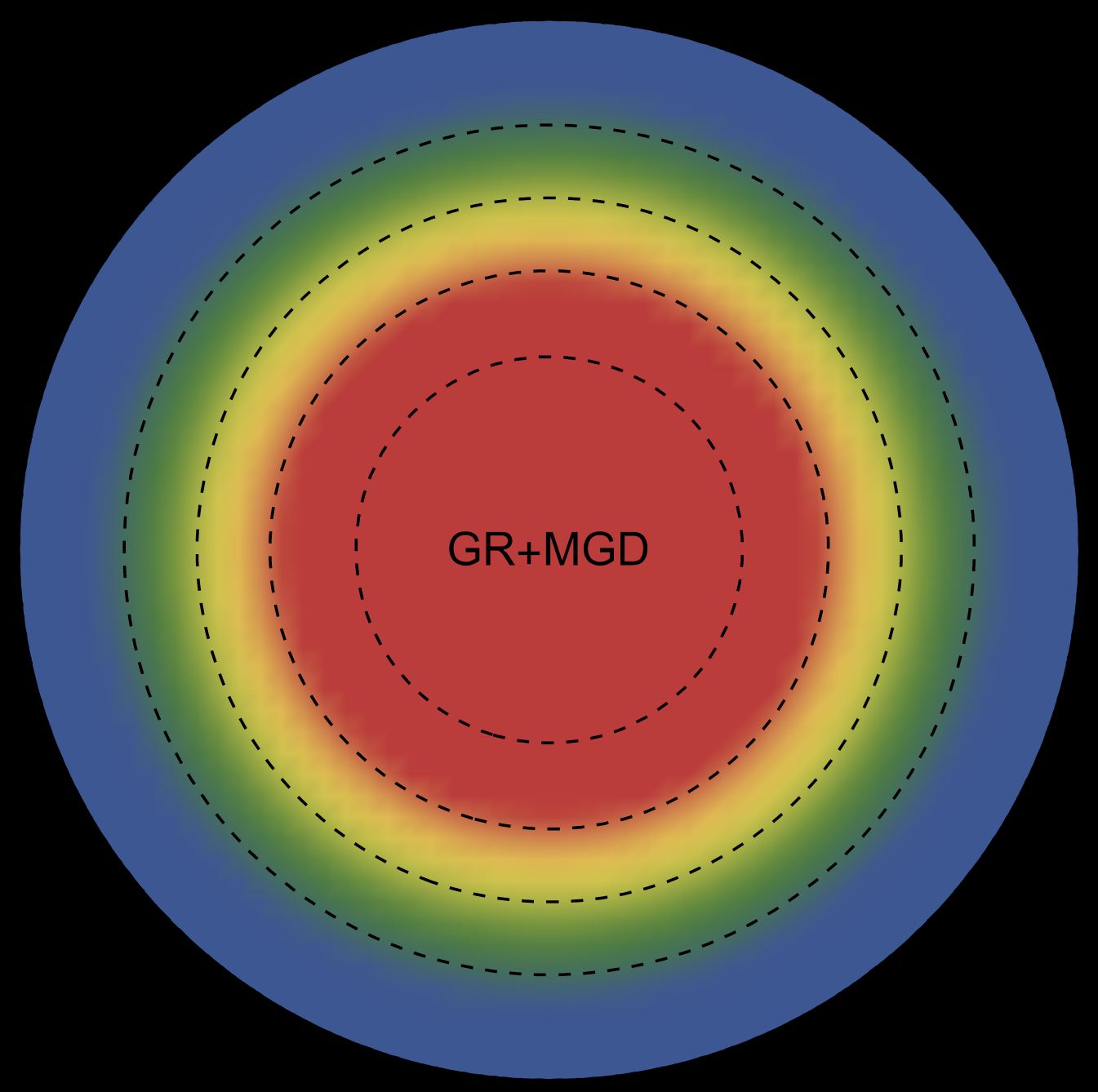}~~\includegraphics[height=5.5cm,width=1.5cm]{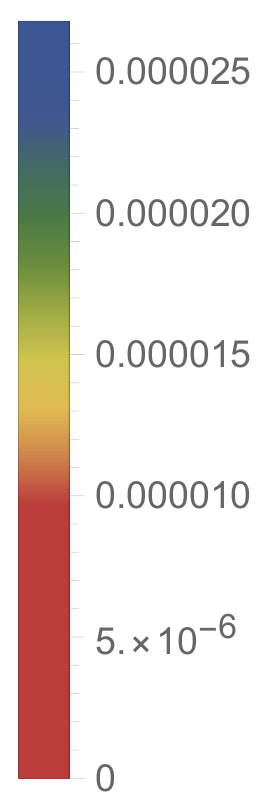}
    \caption{Anisotropy ($\Delta^{\text{eff}} (r)$) profile  within the stellar object corresponding to GR ($\beta=0$) and GR+MGD ($\beta=0.1$) for the parameter values $\epsilon_0=0.0003~\text{km}^{-2}$, $\epsilon_s= 0.00024~\text{km}^{-2}$ and $r_s=11.5~\text{km}$ for solution~\ref{solA} $(\epsilon=\theta^0_0)$.}
    \label{fig4a}
\end{figure}

\begin{figure}[!htp]
    \centering
\includegraphics[height=5.8cm,width=6.0cm]{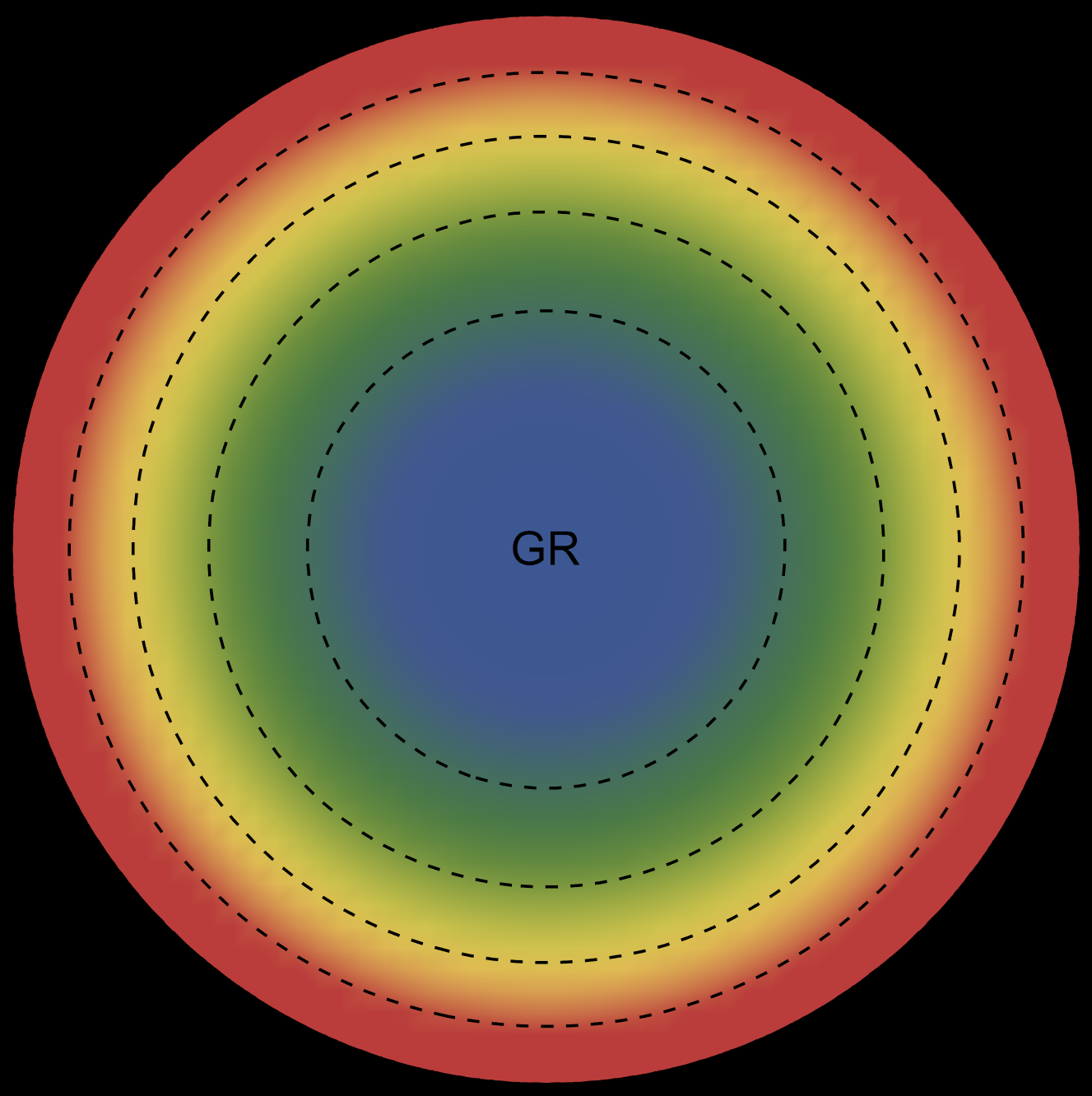}~~\includegraphics[height=5.5cm,width=1.5cm]{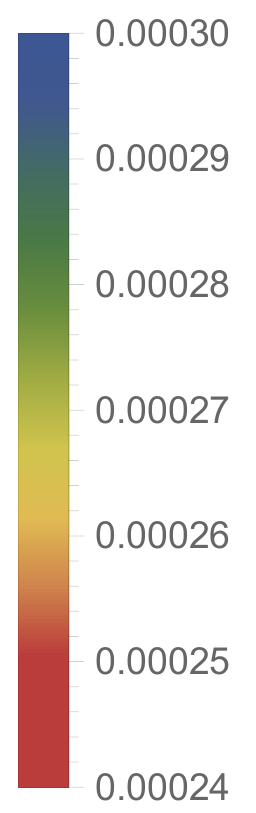}~~~~~~\includegraphics[height=5.8cm,width=6.0cm]{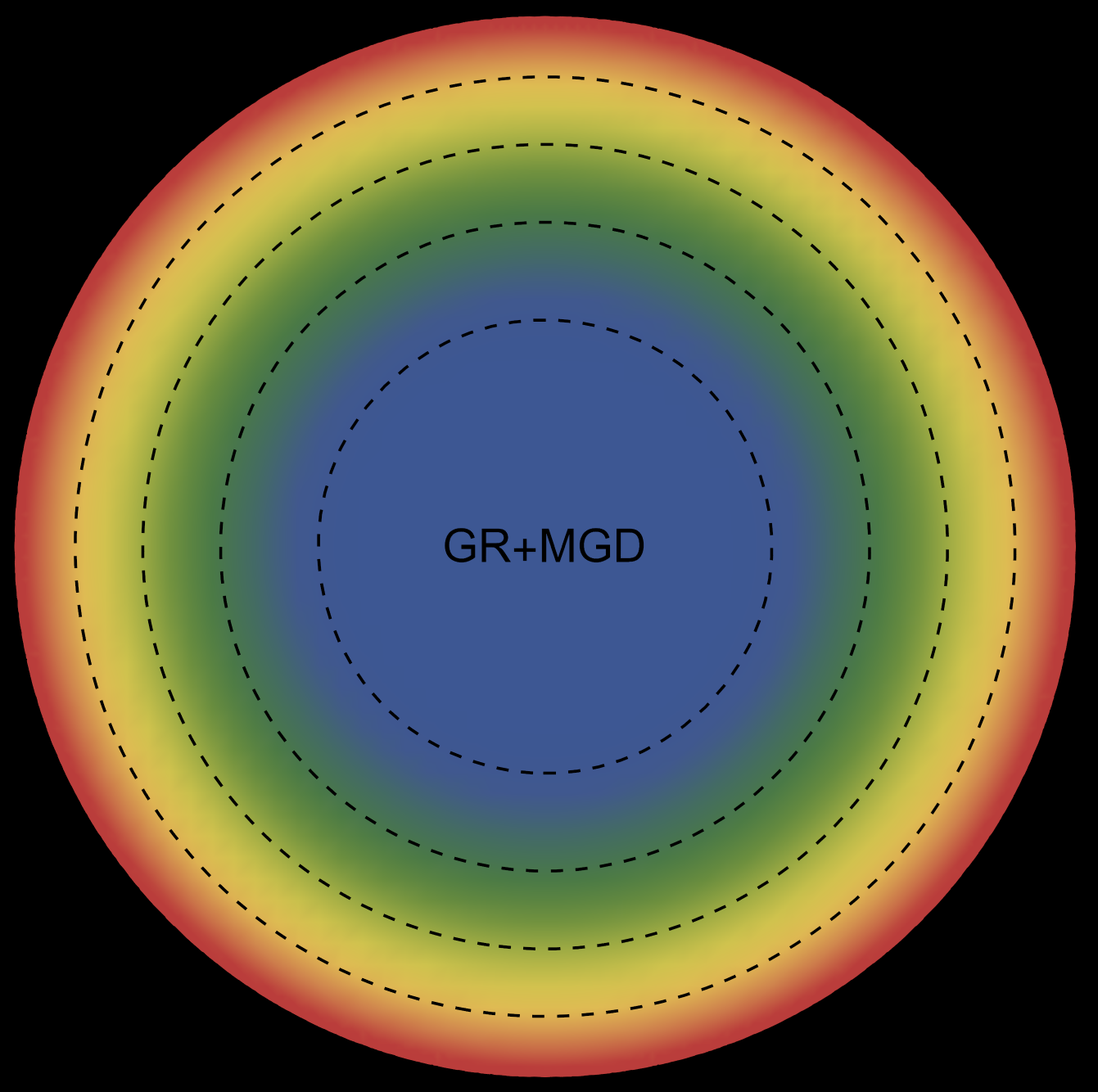}~~\includegraphics[height=5.5cm,width=1.5cm]{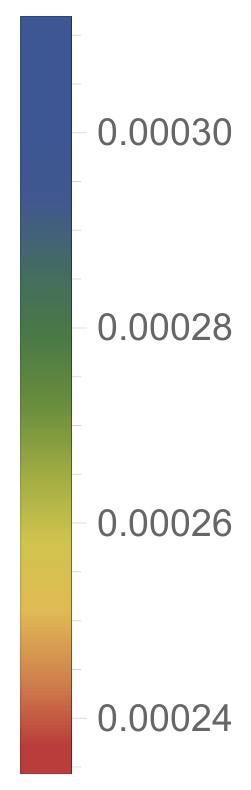}
    \caption{Density ($\rho^{\text{eff}} (r)$) profile  within the stellar object corresponding to GR ($\beta=0$) and GR+MGD ($\beta=0.1$) for the parameter values $\epsilon_0=0.0003~\text{km}^{-2}$, $\epsilon_s= 0.00024~\text{km}^{-2}$ and $r_s=11.5~\text{km}$ for solution~\ref{solB} $(P_r=\theta^1_1)$.}
    \label{fig1b}
\end{figure}

\begin{figure}[!htp]
    \centering
\includegraphics[height=5.8cm,width=6.0cm]{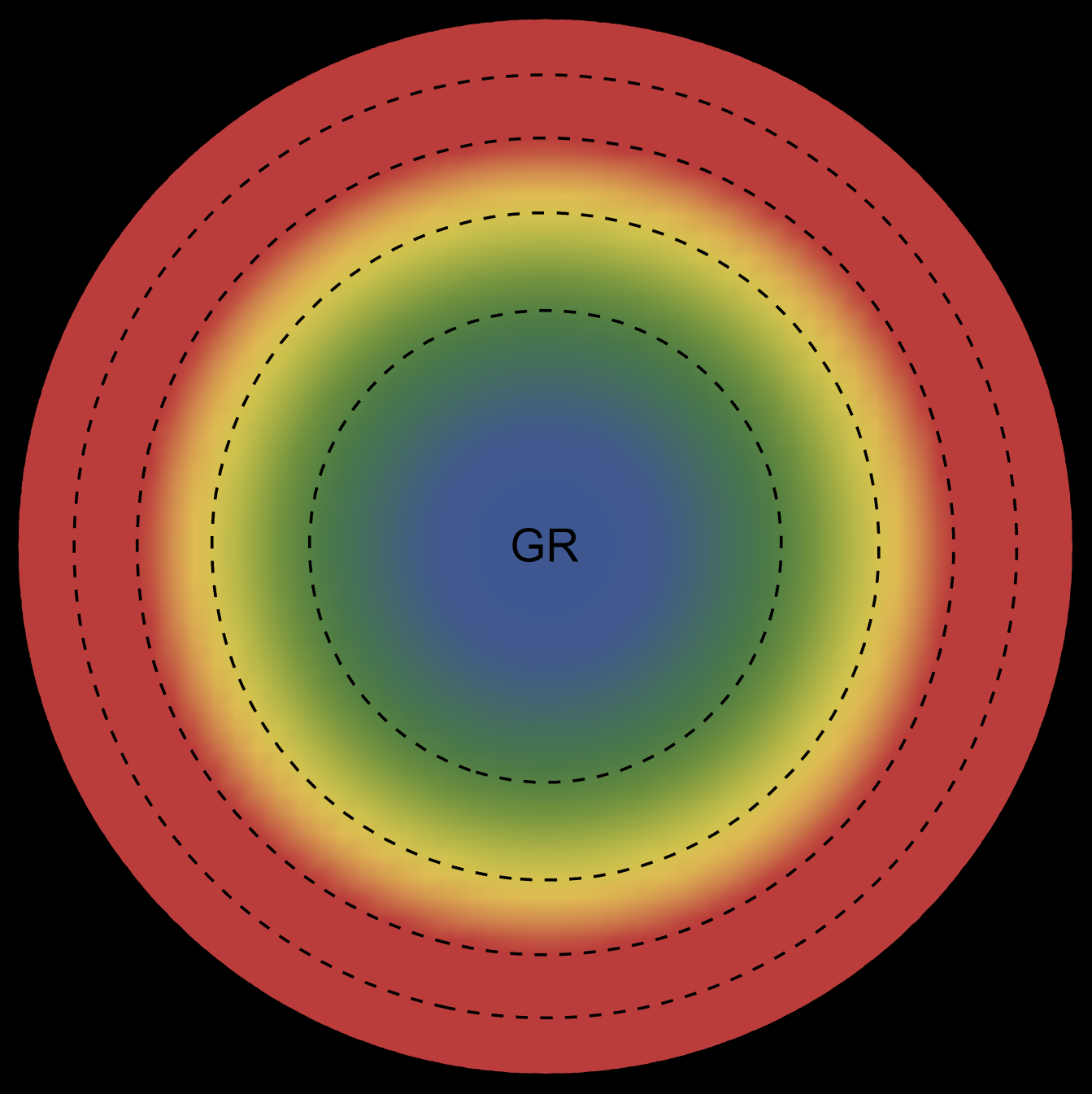}~~\includegraphics[height=5.5cm,width=1.5cm]{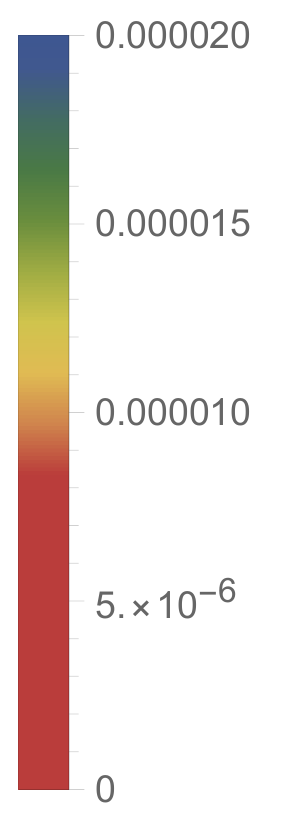}~~~~~~\includegraphics[height=5.8cm,width=6.0cm]{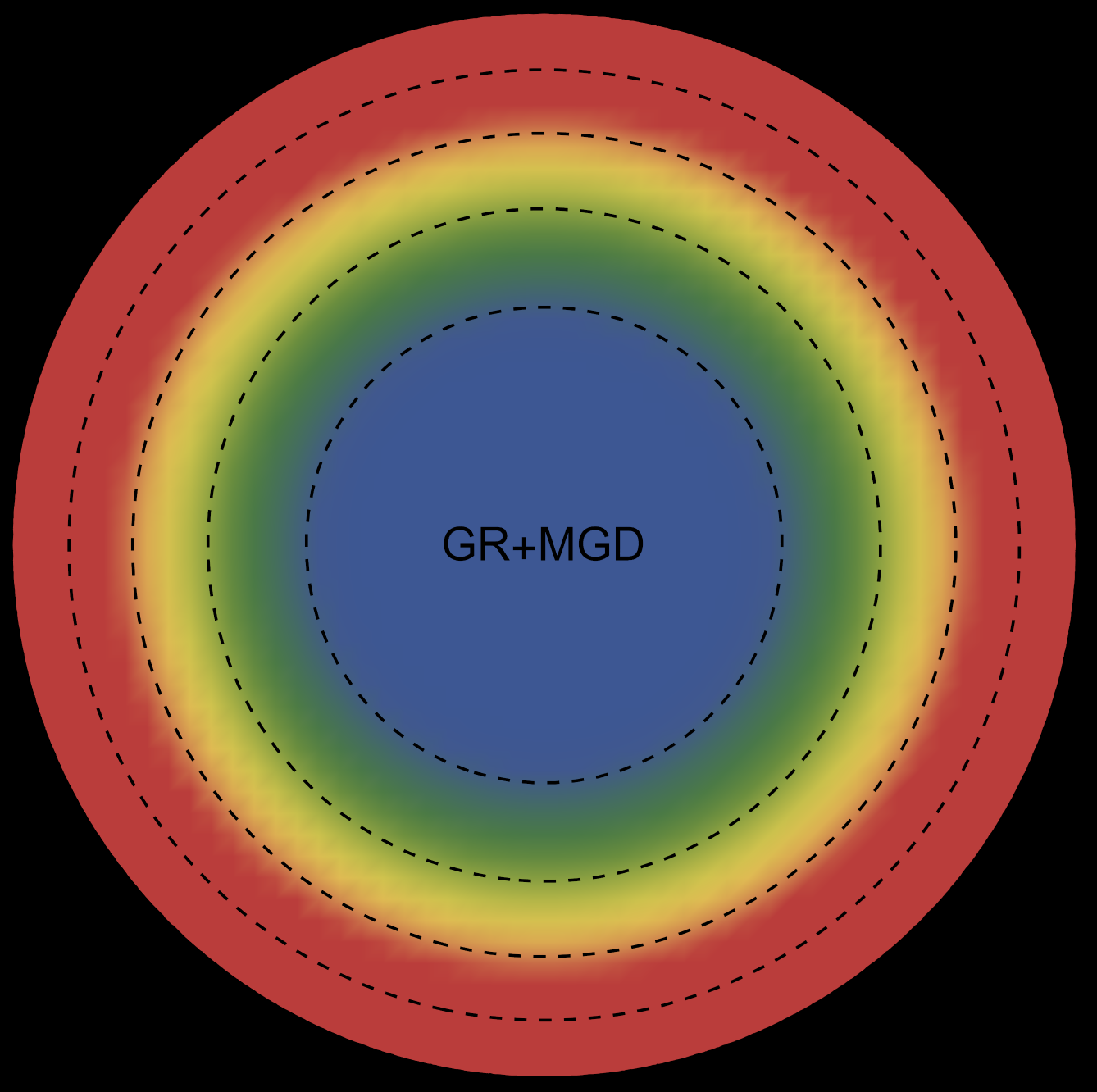}~~\includegraphics[height=5.5cm,width=1.5cm]{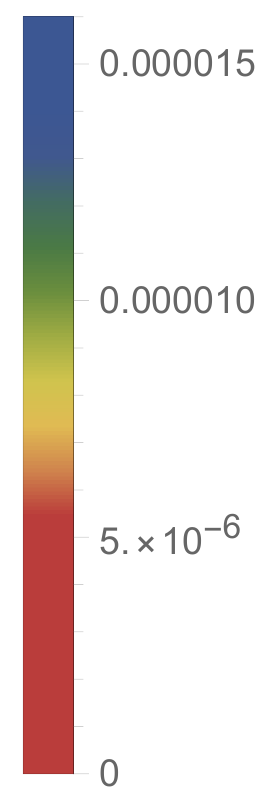}
    \caption{Radial pressure ($P^{\text{eff}}_r (r)$) profile  within the stellar object corresponding to GR ($\beta=0$) and GR+MGD ($\beta=0.1$) for the parameter values $\epsilon_0=0.0003~\text{km}^{-2}$, $\epsilon_s= 0.00024~\text{km}^{-2}$ and $r_s=11.5~\text{km}$ for solution~\ref{solB} $(P_r=\theta^1_1)$.}
    \label{fig2b}
\end{figure}

\begin{figure}[!htp]
    \centering
\includegraphics[height=5.8cm,width=6.0cm]{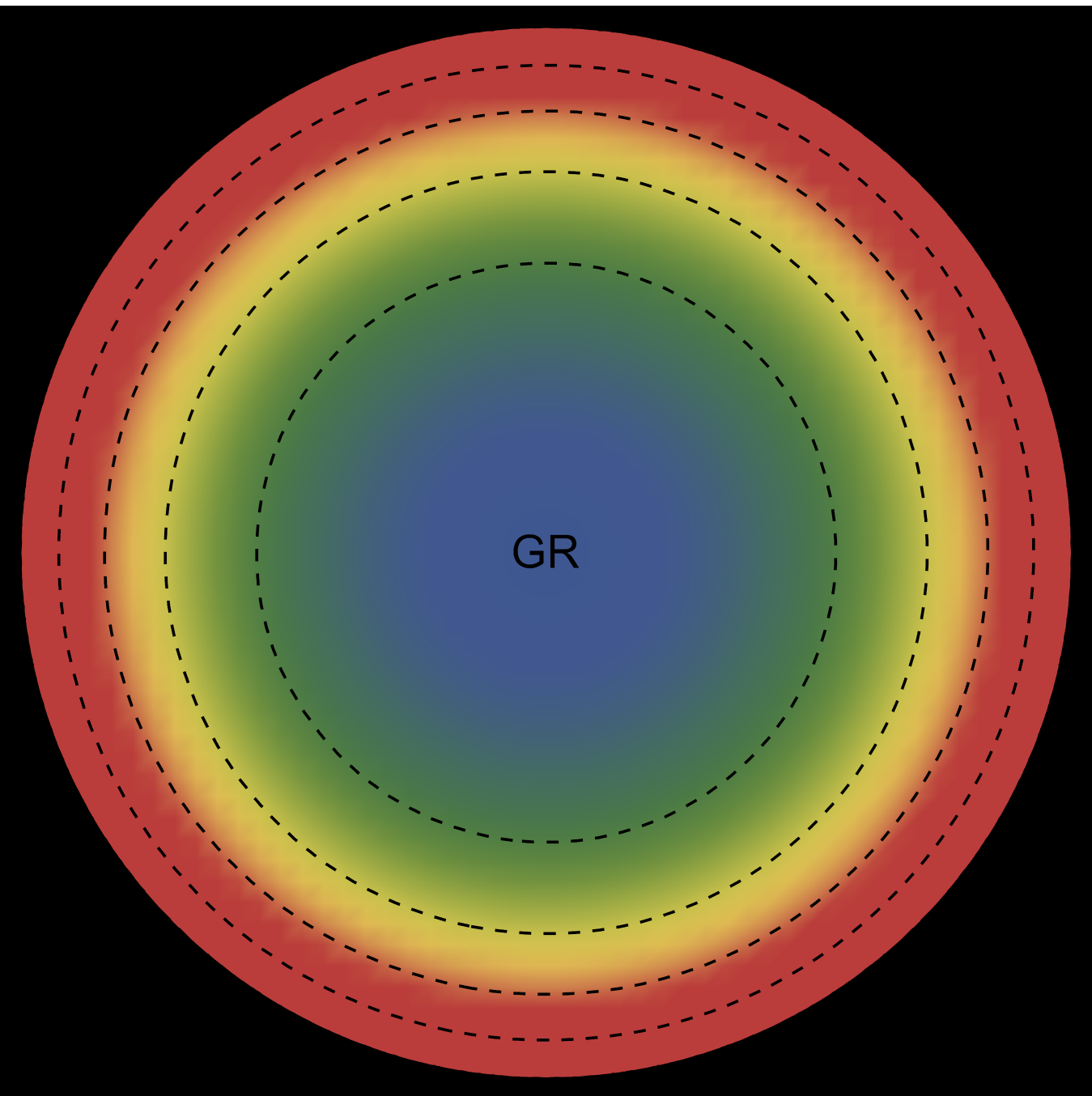}~~\includegraphics[height=5.5cm,width=1.5cm]{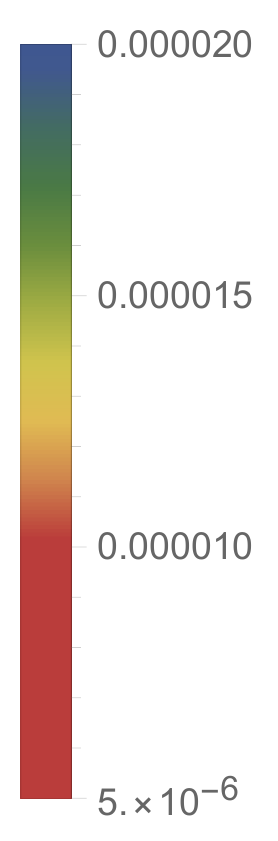}~~~~~~\includegraphics[height=5.8cm,width=6.0cm]{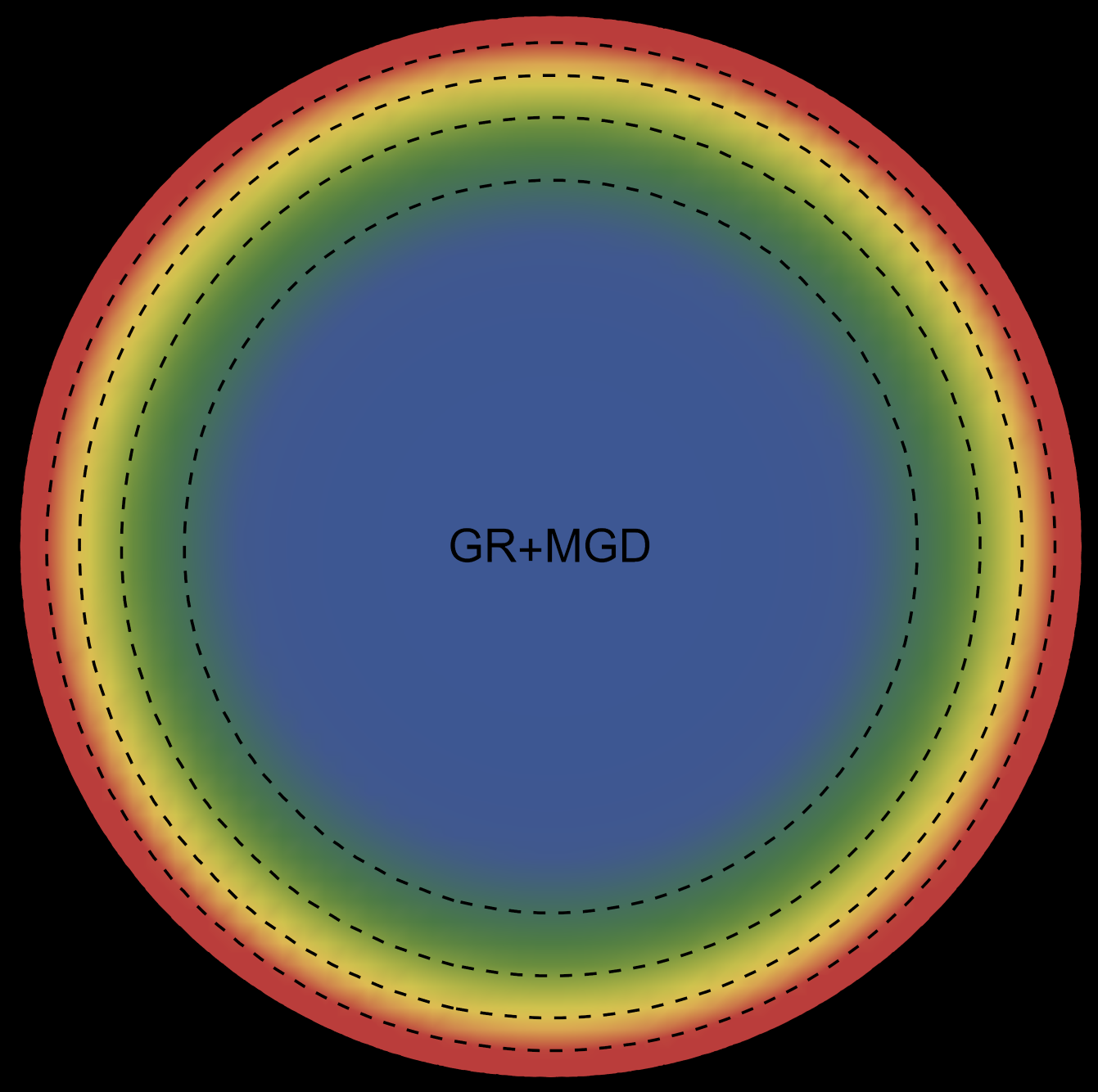}~~\includegraphics[height=5.5cm,width=1.5cm]{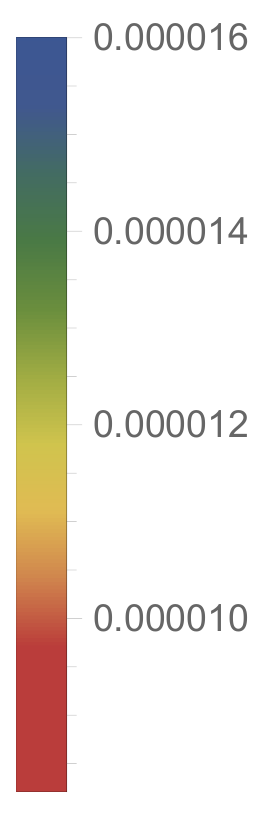}
    \caption{Tangential Pressure ($P^{\text{eff}}_t (r)$) profile  within the stellar object corresponding to GR ($\beta=0$) and GR+MGD ($\beta=0.1$) for the parameter values $\epsilon_0=0.0003~\text{km}^{-2}$, $\epsilon_s= 0.00024~\text{km}^{-2}$ and $r_s=11.5~\text{km}$ for solution~\ref{solB} $(P_r=\theta^1_1)$.}
    \label{fig3b}
\end{figure}

\begin{figure}[!htp]
    \centering
\includegraphics[height=5.8cm,width=6.0cm]{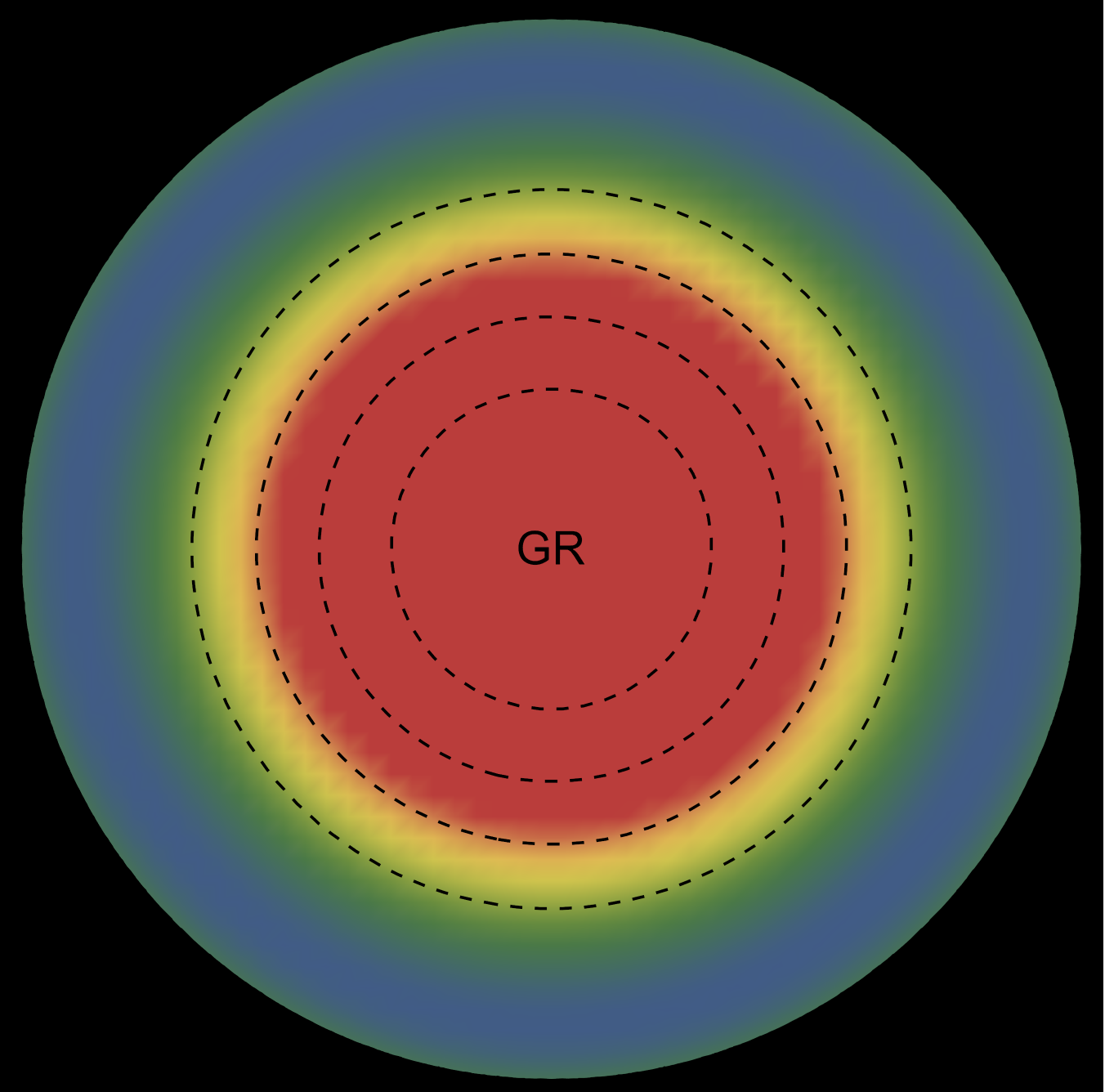}~~\includegraphics[height=5.5cm,width=1.5cm]{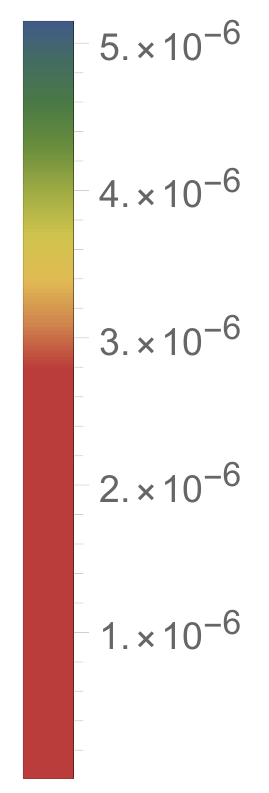}~~~~~~\includegraphics[height=5.8cm,width=6.0cm]{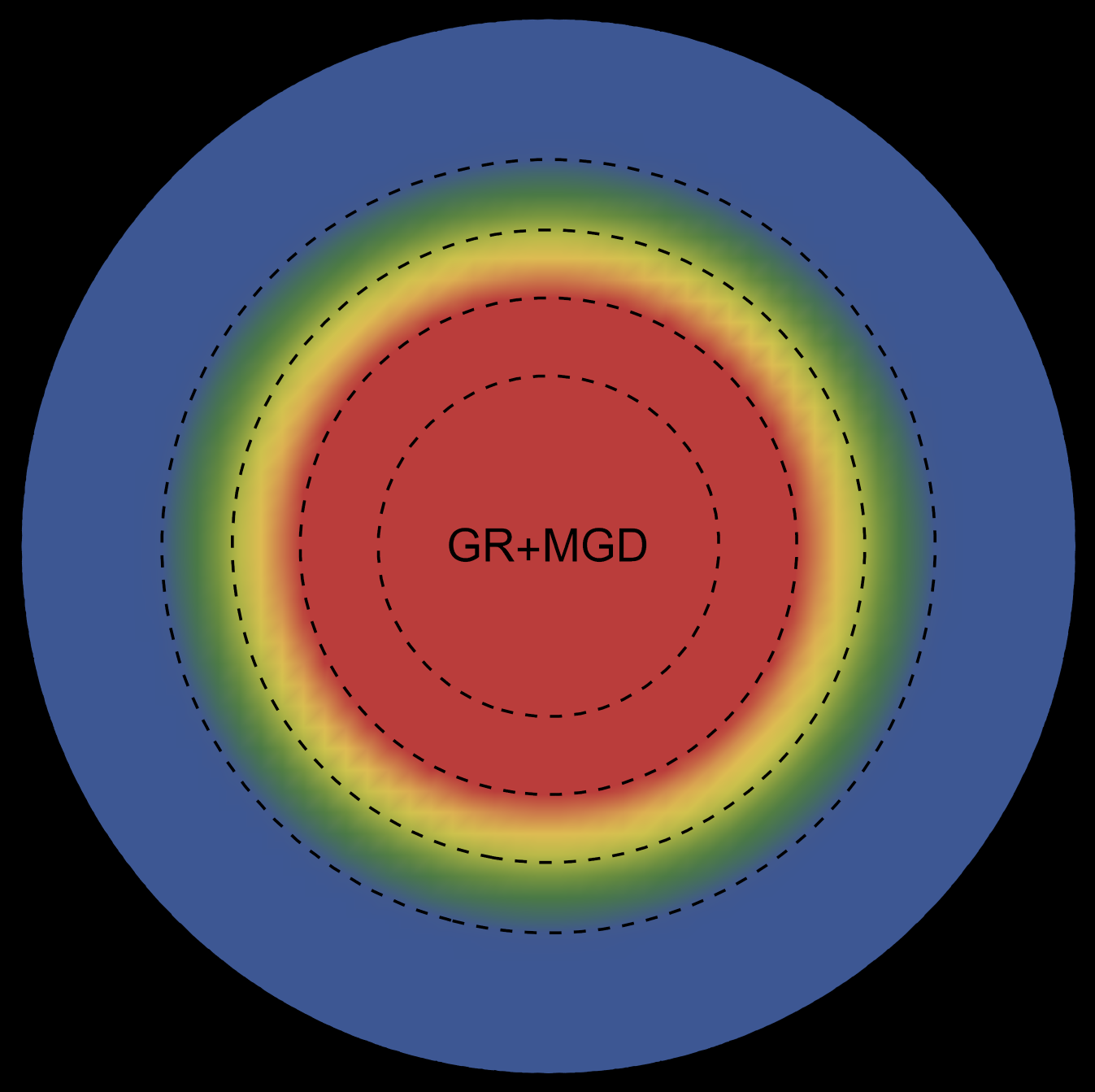}~~\includegraphics[height=5.5cm,width=1.5cm]{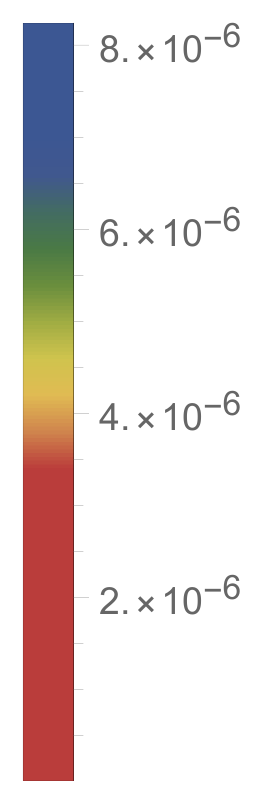}
    \caption{Anisotropy ($\Delta^{\text{eff}} (r)$) profile  within the stellar object corresponding to GR ($\beta=0$) and GR+MGD ($\beta=0.1$) for the parameter values $\epsilon_0=0.0003~\text{km}^{-2}$, $\epsilon_s= 0.00024~\text{km}^{-2}$ and $r_s=11.5$~km for solution~\ref{solB} $(P_r=\theta^1_1)$.}
    \label{fig4b}
\end{figure}

\section{Physical analysis of minimally deformed solutions in strange star models}\label{sec5} 

In this section, we shall examine the physical viability of our deformed strange star models and their significance within astrophysics. Specifically, we aim to analyze the behaviors of various thermodynamic variables, including energy density, radial and tangential stresses, and the anisotropic parameter, along with the mass-radius and mass-moment of inertia profiles. By investigating these elements, we seek to evaluate the relevance of these models in explaining astrophysical phenomena, particularly for the two solutions: the density constraint mimicking $\epsilon(r) = \theta^0_0(r)$~[\ref{solA}] and the pressure constraint mimicking $P_r(r) = \theta^1_1(r)$~[\ref{solB}].

\subsection{Physical behavior of energy density and pressure profiles}

The energy density, $\epsilon(r)$, exhibits an interesting pattern, being largest at the center and consistently decreasing with increasing radial coordinates, reaching its minimum value at the surface of the deformed strange star models. This trend is illustrated in Figs. \ref{fig1a} and \ref{fig1b} for the two solutions. Both figures demonstrate that the energy density remains regular at all interior points for both scenarios: GR (left panel, $\beta=0$) and GR+MGD (right panel, $\beta=0.1$). The parameter values used are $\epsilon_0 = 0.0003 \text{ km}^{-2}$, $\epsilon_s = 0.00024 \text{ km}^{-2}$, and $r_s = 11.5~\text{km}$ for solution~\ref{solA} ($\epsilon = \theta^0_0$) and solution~\ref{solB} ($P_r = \theta^1_1$). In the GR scenario, we observe a lower core density, whereas in the GR+MGD scenario, the core density is elevated. The introduction of MGD results in a higher density in the star's central regions, causing an expansion of matter into concentric shells. However, as one moves from the center towards the surface layers of the star, variations in MGD have minimal impact on the stellar density. It should be noted that raising the parameter $\beta$ enhances the gravitational condensation of matter at lower densities, causing the matter density to shift to lower equilibrium values throughout the stellar interior.

In Figs. \ref{fig2a} and \ref{fig2b} along with \ref{fig3a} and \ref{fig3b}, we show how radial and tangential pressures vary with radial coordinates. These visuals facilitate a comparison of the pressure distributions between the two solutions in each scenario. In both solutions, it is clear that the radial and tangential pressures in the central region are elevated in the GR+MGD scenario compared to the GR scenario while maintaining similar trends and magnitudes. In addition, the integration of the MGD results in elevated radial and tangential pressures in the central region, even though their magnitudes are lower. This observation indicates that the fluid particles are influenced by MGD, leading to a confining and compacting effect, especially in the star's core regions. Crucially, in both solutions, the radial and tangential pressures are continuous across the star and show a consistent decrease as the radial coordinates increase. Notably, the radial pressure component vanishes at the stellar surface, marking an important feature of these solutions.

\subsection{ Physical behavior of anisotropic profiles}

The trend of the anisotropy parameter, $\Delta(r)$, is illustrated in Figs. \ref{fig4a} and \ref{fig4b}, which depict two distinct scenarios that maintain identical fixed parameters. Interestingly, the anisotropy begins at zero at the star's center and increases in magnitude as one moves toward the outer boundary. This increase corresponds to a positive value of $\Delta(r)$, suggesting that radial stresses are stronger than transverse stresses within the star, leading to a repulsive force. This growing anisotropy, along with the resultant repulsive force, is crucial for the stability of the star's surface layers. By resisting the inward gravitational force, these factors contribute to the overall stability of the star. Consequently, the interaction between the growing anisotropy, the repulsive forces, and the equilibrium with gravitational forces is vital for preserving the stability of the star's surface and reinforcing the star's resilience. It is worth emphasizing that, in contrast to GR scenarios, the inclusion of MGD results in a significant increase in anisotropy within the star. Notably, our findings indicate that the anisotropy parameter doubles when MGD is considered, compared to cases where only GR is applied. This observation highlights the profound effect of MGD on the anisotropic behavior of the system, demonstrating that it substantially elevates the anisotropy within the star. The enhancement of anisotropy driven by MGD results from a combination of intricate physical processes occurring inside the star. The transport of neutrinos and electrons is particularly influential, as it alters the pressure distribution and leads to varying levels of anisotropy throughout the stellar structure. Concurrently, phase transitions can change the EOS, impacting the equilibrium between radial and tangential stresses. Furthermore, dissipative effects, including viscosity and heat conduction, also play a significant role in the emergence of anisotropic stresses. Collectively, these interactions create a dynamic and complex environment that markedly increases anisotropy within the star.

\subsection{Physical behavior of $M-R$ and $M-I$ profiles}

\begin{figure}[!htp]
    \centering
\includegraphics[height=6.5cm,width=7.5cm]{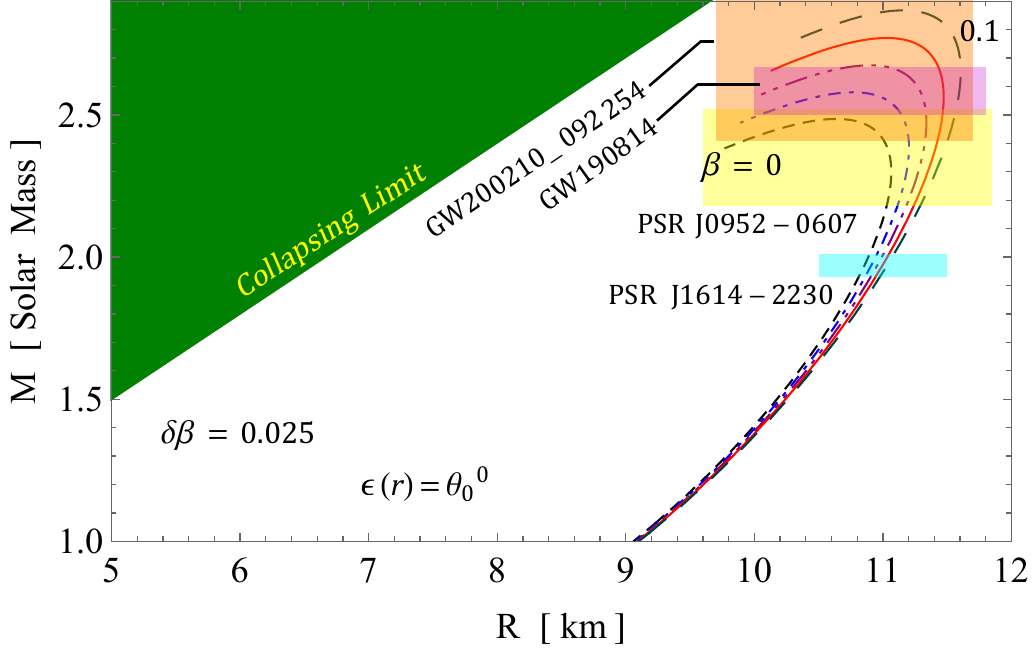}~~~~~~\includegraphics[height=6.5cm,width=7.5cm]{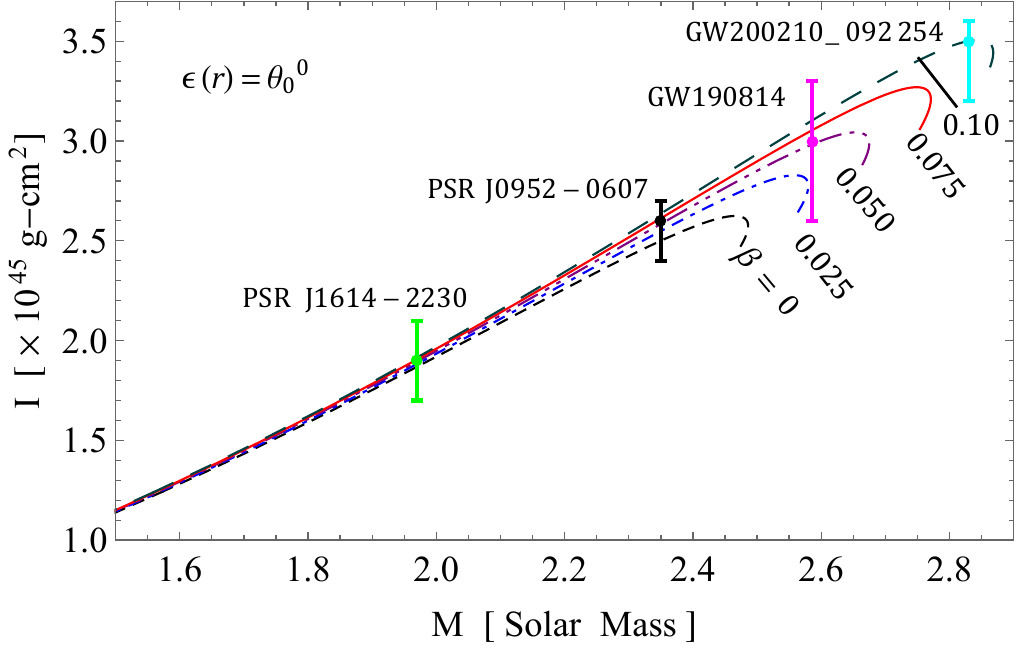}\\
\includegraphics[height=6.5cm,width=7.5cm]{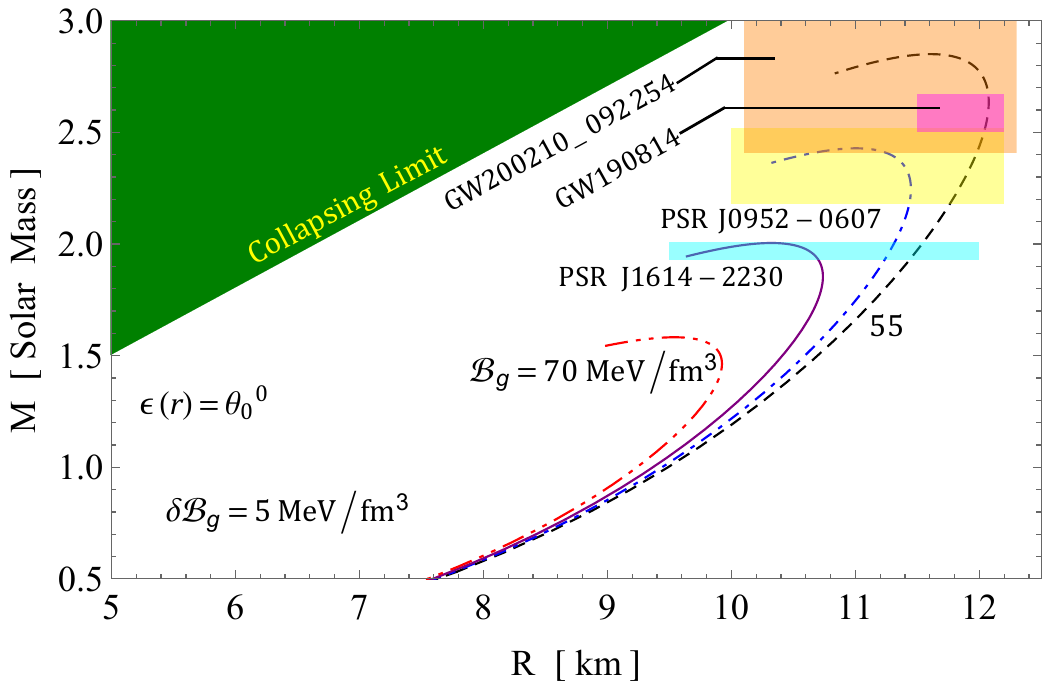}~~~~~~~\includegraphics[height=6.5cm,width=7.5cm]{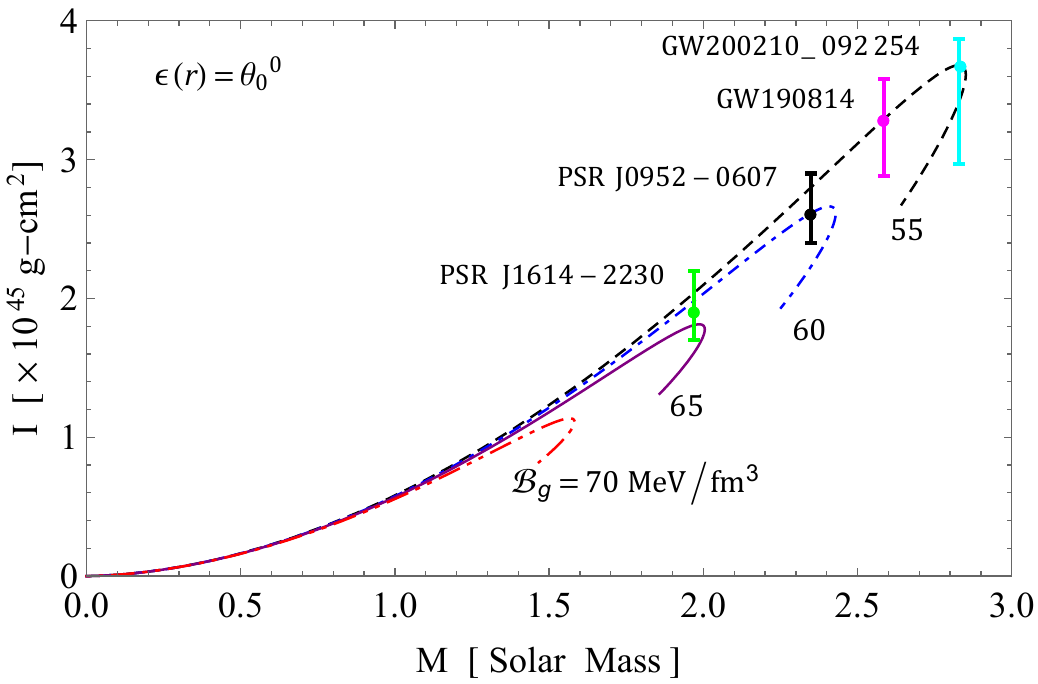}
    \caption{$M-R$ and $M-I$ curves for different values of $\beta$ and $\mathcal{B}_g$.}
    \label{fig5a}
\end{figure}

\begin{figure}[!htp]
    \centering
\includegraphics[height=6.5cm,width=7.5cm]{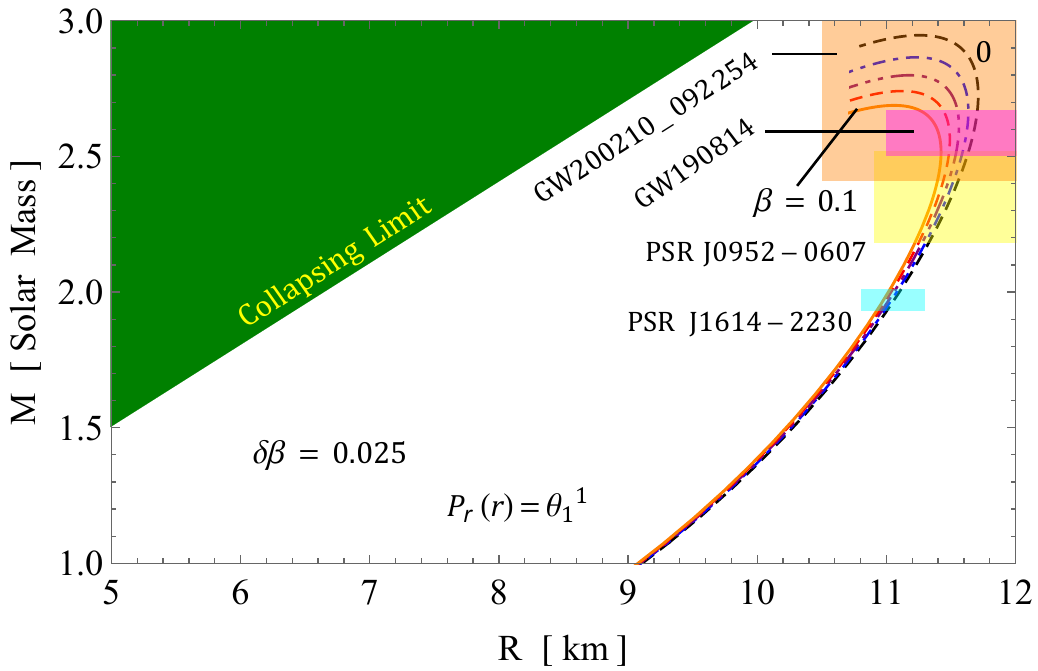}~~~~~~\includegraphics[height=6.5cm,width=7.5cm]{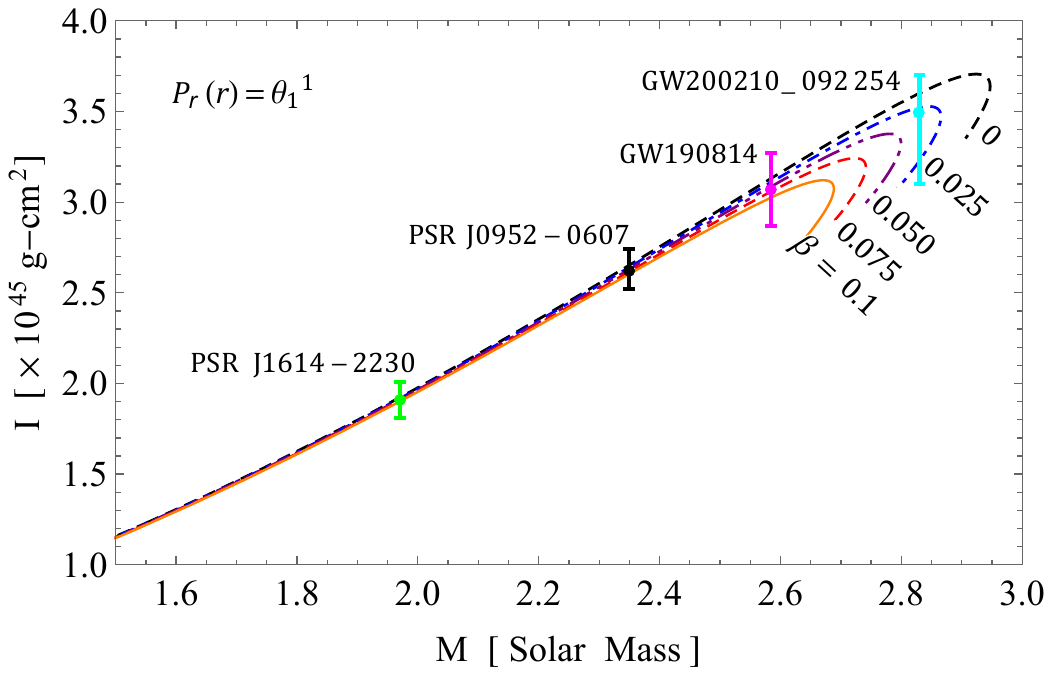}\\
\includegraphics[height=6.5cm,width=7.5cm]{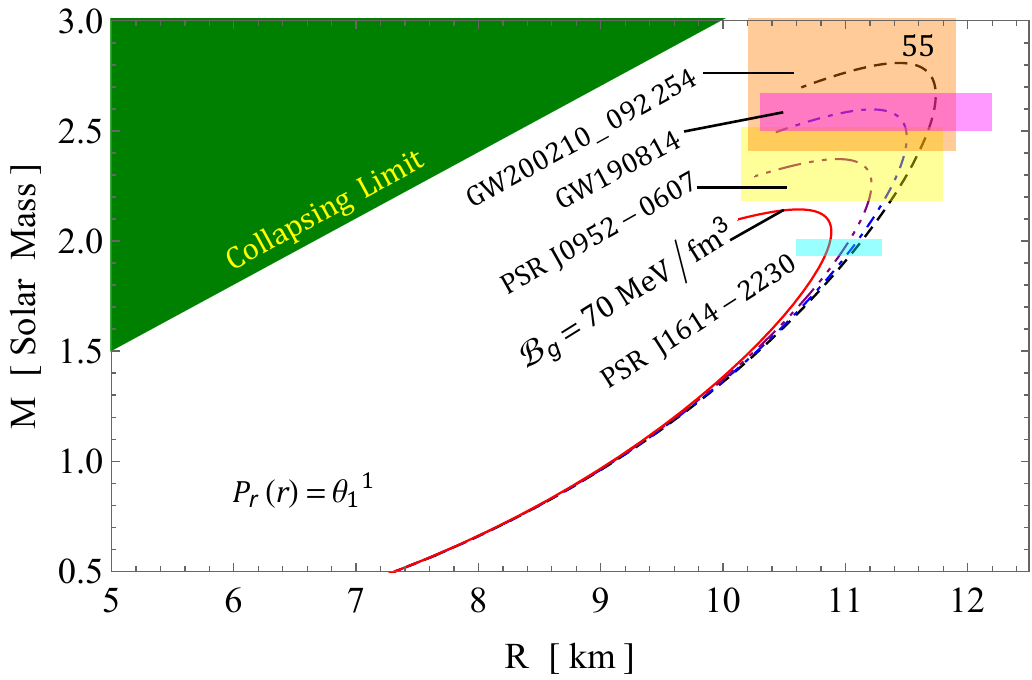}~~~~~~~\includegraphics[height=6.5cm,width=7.5cm]{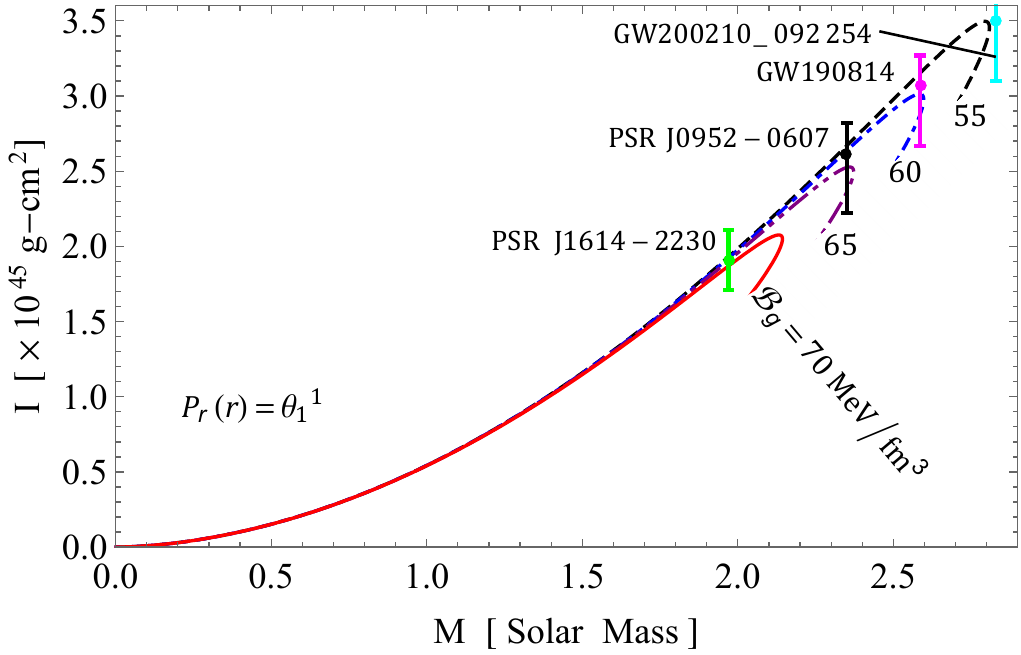}
    \caption{$M-R$ and $M-I$ curves for different values of $\beta$ and $\mathcal{B}_g$.}
    \label{fig5b}
\end{figure}

The $M-R$ and $M-I$ curves are illustrated in Figs. \ref{fig5a} and \ref{fig5b}, respectively, showcasing how they vary with different values of the decoupling constant $\beta$ and the bag constant $\mathcal{B}_g$. These curves provide insight into the behavior of two distinct solutions under consideration:
\begin{itemize}
    \item The first solution [\ref{solA}], employs a density constraint that is represented by $\epsilon(r) = \theta^0_0(r)$. This indicates that the energy density at a specific radius $r$ is directly related to the component of the stress-energy tensor, which captures the distribution of energy within the system.
    \item The second solution [\ref{solB}], relies on a pressure constraint expressed as $P_r(r) = \theta^1_1(r)$. In this case, the radial pressure at the same radius $r$ corresponds to the relevant component of the stress-energy tensor, which describes how pressure is distributed in the system.
\end{itemize}

By analyzing these curves, we gain important insight into how variations in the decoupling constant and bag constant influence the properties of the solutions, providing valuable insight into the underlying physical phenomena. 

The moment of inertia $I$ can be determine via the Bejger and Haensel formula \cite{MI} given as
\begin{eqnarray}
I = \frac{2}{9} \Bigg( 1+\frac{5(M_{NR}/R_{NR}).km }{M_{\odot}} M_{NR} R_{NR}^2\Bigg). \label{e41}
\end{eqnarray}

Here $NR$ represents non-rotating. Notably, the $M-R$ and $M-I$ curves are derived under the assumption that the distribution of SQM among the peculiar stars is governed by the straightforward physical MIT bag model EOS. This is achieved by employing a modified version of matter density proposed by Mak and Harko \cite{Harko:2002pxr}, in conjunction with the TOV equation. This approach enables us to investigate the relationship between mass, radius, and moment of inertia for pulsars in both GR and GR+MGD scenarios. We have derived estimates for the radii and moments of inertia of four different massive stellar objects--PSR J1614-2230 \citep{Demorest:2010bx}, PSR J0952-0607 \citep{Romani:2022jhd}, GW190814 \citep{LIGOScientific:2020zkf}, and GW200210 \citep{KAGRA:2021vkt}--based on the $M-R$ and $M-I$ curves corresponding to their observed masses. We have investigated key findings on the maximum mass as indicated by the $M-R$ and $M-I$ curves. 
\begin{itemize}
    \item In solution \ref{solA}, as the parameter $\beta$ increases from 0 to 0.1 and $\mathcal{B}_g$ decreases from 70 \, MeV/fm$^3$ to 55 \, MeV/fm$^3$, we observe from Fig. \ref{fig5a} that the maximum mass significantly rises, peaking at $\beta = 0.1$ and $\mathcal{B}_g = 55 \, \text{MeV/fm}^3$. Furthermore, the figure illustrates that for all increasing values of $\beta$ and decreasing values of $\mathcal{B}_g$, the maximum moment of inertia $I$ initially rises from zero to a certain mass limit before rapidly declining. This leads us to conclude that the sensitivity of the $M-I$ curves is enhanced, indicating that the rigidity of the EOS improves when $\beta$ is increased and $\mathcal{B}_g$ is decreased. Tables \ref{tab1} and \ref{tab2} reveal that the predicted radii and moments of inertia for the stellar objects exhibit a notable increase as the parameter $\beta$ rises and the parameter $\mathcal{B}_g$ decreases. According to Table \ref{tab5}, the maximum gravitational mass reaches $M_{\text{max}} = 2.87 \, M_\odot$ with a radius of 11.20 km for $\beta = 0.1$ and $I_{max}=3.44\times 10^{45}\,g\cdot cm^2$. In contrast, for $\beta = 0$, we observed $M_{\text{max}} = 2.48 \, M_\odot$ with a radius of 10.69 km and $I_{max}=2.58\times 10^{45}\,g\cdot cm^2$. Additionally, for $\mathcal{B}_g = 55 \, \text{MeV/fm}^3$, the maximum gravitational mass is $M_{\text{max}} = 2.85 \, M_\odot$ with a radius of 11.68 km and $I_{max}=3.61\times 10^{45}\,g\cdot cm^2$, while for $\mathcal{B}_g = 70 \, \text{MeV/fm}^3$, it is recorded as $M_{\text{max}} = 1.58 \, M_\odot$ with a radius of 9.64 km and $I_{max}=1.10\times 10^{45}\,g\cdot cm^2$. Notably, all sequences of compact objects achieve a maximum mass of at least $M_{\text{max}} > 2 \, M_\odot$, except in the case of $\mathcal{B}_g = 70 \, \text{MeV/fm}^3$. Our calculations indicate that when $\beta \neq 0$, the maximum mass of the corresponding compact object exceeds the lower mass limit of the secondary components of GW190814 and GW200210. This trend suggests a significant correlation between these parameters and the physical characteristics of the stellar objects, highlighting how variations in $\beta$ and $\mathcal{B}_g$ influence their structural properties.
    \item In Solution \ref{solB}, we observe a contrasting behavior for the parameter $\beta$, while the trend for $\mathcal{B}_g$ remains consistent, albeit with a slight change in magnitude. Specifically, as $\beta$ decreases from 0.1 to 0 and $\mathcal{B}_g$ decreases from $70 \, \text{MeV/fm}^3$ to $55 \, \text{MeV/fm}^3$, Fig. \ref{fig5b} shows a significant increase in the maximum mass, which peaks at $\beta = 0$ and $\mathcal{B}_g = 55 \, \text{MeV/fm}^3$. Additionally, the figure indicates that for all decreasing values of $\beta$ and $\mathcal{B}_g$, the maximum moment of inertia $I$ initially increases from zero until it reaches a certain mass limit, after which it declines rapidly. This observation further suggests that the sensitivity of the $M-I$ curves is heightened, reflecting an improvement in the rigidity of the EOS as both $\beta$ and $\mathcal{B}_g$ decrease. Moreover, Tables \ref{tab3} and \ref{tab4} demonstrate that the predicted radii and moments of inertia for the stellar objects significantly increase with decreasing values of $\beta$ and $\mathcal{B}_g$. It is clearly illustrated in Table \ref{tab5} that the maximum gravitational mass reaches $M_{\text{max}} = 2.95 \, M_\odot$ ($I_{max}=3.64\times 10^{45}\,g\cdot cm^2$) with a radius of 11.32 km for $\beta = 0$ and $\mathcal{B}_g = 55 \, \text{MeV/fm}^3$. In comparison, for $\beta = 0.1$ and $\mathcal{B}_g = 70 \, \text{MeV/fm}^3$, we observe $M_{\text{max}} = 2.69 \, M_\odot$ ($I_{max}=3.02\times 10^{45}\,g\cdot cm^2$) and $M_{\text{max}} = 2.14 \, M_\odot$ ($I_{max}=1.99\times 10^{45}\,g\cdot cm^2$) with radii of 11.09 km and 10.67 km for $\beta = 0$ and $\mathcal{B}_g = 55 \, \text{MeV/fm}^3$, respectively. Similarly, in the initial scenario, we find that all sequences of compact objects achieve a maximum mass of at least $M_{\text{max}} > 2 \, M_\odot$ in all cases. Therefore, it is important to reiterate that for all values of $\beta$ chosen, the maximum mass of the corresponding compact object surpasses the lower mass limit of the secondary components of GW190814 and GW200210. This trend indicates a strong correlation between these parameters and the characteristics of the stellar objects, showing how variations in $\beta$ and $\mathcal{B}_g$ influence their structure.
\end{itemize}

In recent studies exploring the properties of nucleonic matter through variational methods and EOSs, researchers have observed intriguing $M-R$ curves that consistently indicate a substantial mass near $3 M_\odot$. This observation parallels the $M-R$ curves illustrated in Figs \ref{fig5a}, and \ref{fig5b}. One significant contribution to this field is the work by \cite{Astashenok:2021peo}, which investigated the $\mathcal{R}^2$ model scenario. Their findings demonstrated that the $M-R$ curve in this framework aligns with the general relativistic limit, approximately at $3 M_{\odot}$. Notably, the causal mass upper limit was found to correspond closely with the general relativistic maximum mass, positioning it within the mass gap. In addition, this research delved into the characteristics of strange stars in the context of modified gravity, specifically relating to the secondary component of the GW190814 event. Another important study by \cite{Astashenok:2020qds} examined the existence of supermassive compact stars, identifying masses ranging from roughly $2.2$ to $2.3 M_{\odot}$ and radii around $11$ km, all framed within the axion $\mathcal{R}^2$ gravity model. Further expanding on these findings, \cite{Astashenok:2020cqq} analyzed rotating NS models using the GM1 EOS and compared their findings with static stars, presenting mass-radius and mass-central density relationships. Additionally, they explored realistic supermassive NSs, finding masses between $2.2$ and $2.3 M_\odot$ and a radius of $11$ km with the APR EOS \cite{Astashenok:2020cfv}. In perturbative $f(\mathcal{R})$-gravity, stable NSs were derived using FPS and SLy EOS, yielding a minimum radius of $9$ km and a maximum mass of $1.9 M_\odot$ \cite{Astashenok:2013vza}. Further, they examined extreme NSs within $f(G)$ and $f(\mathcal{R})$ theories, demonstrating the potential for stars exceeding $4 M_\odot$ with radii between $12$ and $15$ km \cite{Astashenok:2014nua}. Their investigations also included various EOS within different inflationary scenarios, highlighting that MPA1 is the only EOS satisfying all constraints, with masses exceeding $2.5 M_\odot$ but below the $3 M_\odot$ causal limit \cite{Astashenok:2017dpo,Odintsov:2023ypt}. Capozziello and collaborators contributed by discussing anisotropic stars and mass-radius relations in this framework \cite{Astashenok:2021peo,Astashenok:2020qds,Nashed:2021sji}.

\section{Stability analysis}\label{sec6}

\subsection{Stability analysis via adiabatic index} 

We now need to evaluate the stability of the anisotropic stellar configuration. This requires an analysis of the adiabatic index ($\Gamma$), which is defined by the following expression
\begin{equation}
    \Gamma = \frac{\rho+P_r}{P_r}\frac{dP_r}{d\rho}.
\end{equation}

The stability condition for an isotropic fluid in the Newtonian framework is formulated as $\Gamma > \frac{4}{3}$. This criterion is highlighted in the studies by Heintzmann and Bondi \cite{heintzmann1975neutron,bondi1992anisotropic}, which focus on neutron stars and anisotropic fluids. In contrast, the stability requirement for an anisotropic stellar model diverges from the traditional Chandrasekhar result applicable to isotropic fluids \cite{chan1992dynamical,chan1993dynamical}
\begin{equation}
\Gamma > \frac{4}{3}\left(1 + \frac{\Delta}{r|(P_{r})|^{\prime}}+\frac{1}{4}\frac{\kappa \rho P_{r}r}{|(P_{r})|^{\prime}} \right).\label{eq62}
\end{equation}

In this context, primes indicate differentiation with respect to the radial coordinate, denoted as $r$. The second term in Eq. (\ref{eq62}) addresses the modifications to the stability condition that arise from anisotropy ($P_r \neq P_t$), while the last term incorporates relativistic corrections.

The value of the adiabatic index, denoted as $\Gamma$, is subject to specific constraints related to the dynamical instability of an isotropic fluid sphere. This critical threshold is referred to as the critical adiabatic index, $\Gamma_{cr}$, which is defined by the inequality 
\begin{equation}
\langle \Gamma \rangle > \Gamma_{cr},
\end{equation}
where $\langle \Gamma \rangle$ represents the averaged adiabatic index \cite{Moustakidis:2016ndw}. 

According to Refs. \cite{Moustakidis:2016ndw, Koliogiannis:2018hoh}, the critical value can be formulated as 
\begin{equation}
\Gamma_{cr} = \frac{4}{3} + \frac{19}{42} C,
\end{equation}
where $C = \frac{2M}{R}$ signifies the compactness parameter within the framework of General Relativity. In the context of pure Newtonian gravity, the critical value remains constant at $\Gamma_{cr} = \frac{4}{3}$. However, when considering the influences of GR, this value increases beyond $\frac{4}{3}$.

\begin{figure}[!htp]
    \centering
\includegraphics[height=6.8cm,width=7.5cm]{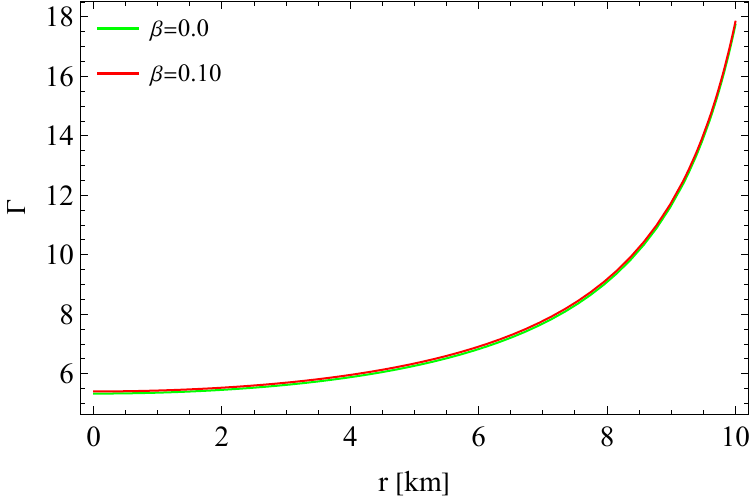}~~~\includegraphics[height=6.8cm,width=7.5cm]{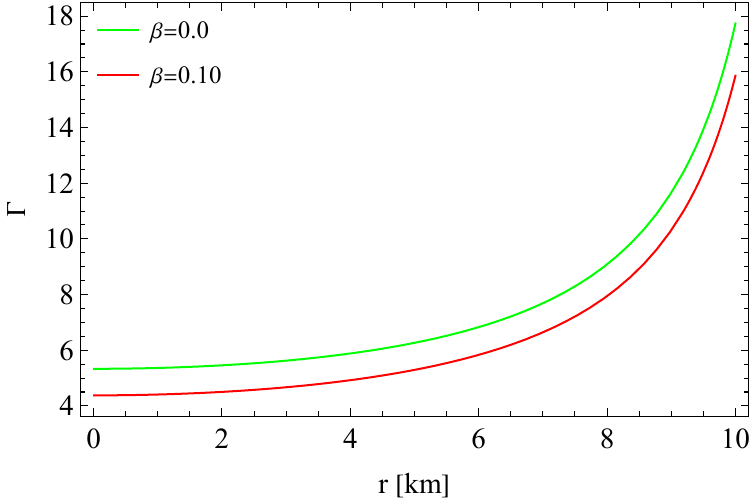}
    \caption{Adiabatic index ($\Gamma$) profile  within the stellar object corresponding to GR ($\beta=0$) and GR+MGD ($\beta=0.1$) for the parameter values $\epsilon_0=0.0003~\text{km}^{-2}$, $\epsilon_s= 0.00024~\text{km}^{-2}$ and $r_s=11.5$~km for solutions~ \ref{solA} (left panel) and \ref{solB} (right panel).}
    \label{fig4c}
\end{figure}

To further investigate the behavior of the adiabatic index $\Gamma$, we refer to Fig. \ref{fig4c}, which utilizes the same parameter sets as those presented in Figs. \ref{fig1a} to \ref{fig1b}. The results obtained reveal a strong correlation with a dynamically stable configuration for the two solutions: the density constraint mimicking $\epsilon(r) = \theta^0_0(r)$~[\ref{solA}] and the pressure constraint mimicking $P_r(r) = \theta^1_1(r)$~[\ref{solB}].

\begin{figure}[!htp]
    \centering
\includegraphics[height=6.8cm,width=7.5cm]{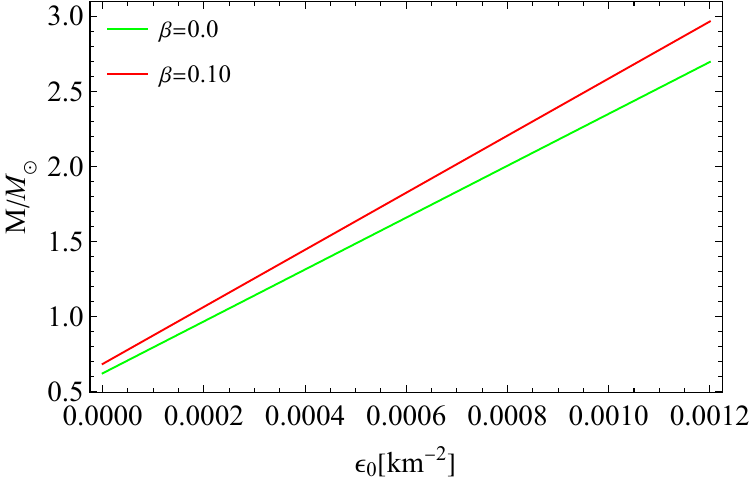}~~~\includegraphics[height=6.8cm,width=7.5cm]{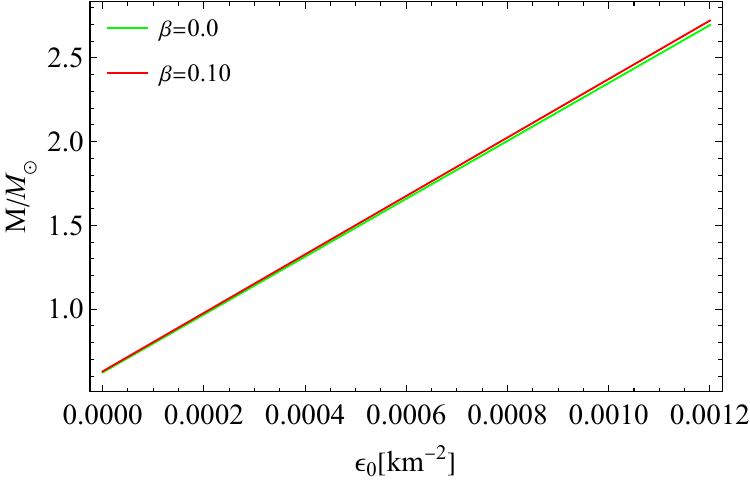}
    \caption{Stability analysis via $M-\rho^{\text{eff}}_0$ within the stellar object corresponding to GR ($\beta=0$) and GR+MGD ($\beta=0.1$) for the parameter values $\epsilon_0=0.0003~\text{km}^{-2}$, $\epsilon_s= 0.00024~ \text{km}^{-2}$ and $r_s=11.5$~km for solutions~\ref{solA} (left panel) and \ref{solB} (right panel).}
    \label{fig4d}
\end{figure}

\subsection{Stability analysis via Harrison-Zel'dovich-Novikov criteria} 
We now explore the relationship between stellar mass $M$ and central energy density $\epsilon_c$ for two solutions: the density constraint given by $\epsilon(r) = \theta^0_0(r)$~[\ref{solA}]  and the pressure constraint given by $P_r(r) = \theta^1_1(r)$~[\ref{solB}]. To achieve this, we apply the static stability criterion \cite{ZHN1, ZHN2}
 \begin{itemize}
\item~~~~ $\frac{dM}{d\,\epsilon_0} < 0 \hspace{0.5cm} \rightarrow \mbox{unstable configuration}$,\\
 \item ~~~~  $\frac{dM}{d\, \epsilon_0} > 0 \hspace{0.5cm} \rightarrow \mbox{stable configuration}$,
\end{itemize}
to be satisfied by all stellar configurations, where $\epsilon_0$ is the central density of the seed system. 

To verify the above-mentioned condition for both solutions, we derive the mass equation as an algebraic function of $\epsilon_0$, expressed in the following formulas:
\begin{itemize}
\item~ $M=\frac{4}{15} \pi  (\beta +1) r_s^3 (2 \epsilon_0+3 \epsilon_s)$,~~~~~~~~~~~~~~~~~~~~~~~~~~~~~~~~~~~~~~~~~~~~~~~~~~~(\text{For solution}~\ref{solA})\\
 \item ~ $M=\frac{4 \pi  r_s^3}{15 \left(3-32 \pi  \mathcal{B}_g r_s^2+8 \pi  \epsilon_s r_s^2\right)} \Big[ 2 \epsilon_0 \left(32 \pi  (\beta -1) \mathcal{B}_g r_s^2-8 \pi  (\beta -1) \epsilon_s r_s^2+3\right)  \nonumber\\ \hspace{0 cm}-3 \left(20 \beta  \mathcal{B}_g-\epsilon_s \left(5 \beta +32 \pi  (\beta -1) \mathcal{B}_g r_s^2+3\right)+8 \pi  (\beta -1) \epsilon_s^2 r_s^2\right)\Big]$. ~~~~~~~(\text{For solution}~\ref{solB}).
\end{itemize}

From Fig. \ref{fig4d}, it is evident that $M(\epsilon_0)$ is positive and increasing, hence $\frac{dM}{d\epsilon_{_0}}$ will be positive. This indicates that the stellar structures developed are stable. Notably, within the same range of $\epsilon_0$ fluctuations, the inclusion of MGD results in a more significant increase in mass, particularly for the solution with the density constraint, $\epsilon(r) = \theta^0_0(r)$~[\ref{solA}]. In contrast, there is a slight change observed in the solution with the pressure constraint, $P_r(r) = \theta^1_1(r)$~[\ref{solB}], suggesting enhanced stability under small density fluctuations.

\section{Concluding remarks}\label{sec7} 

Our research is centered on simulating the configurations of matter associated with mass gap objects, particularly those formed through NS mergers and the evolution of massive pulsars. Observational data indicate that the distribution of BH masses tends to cluster around specific values, such as those recorded in gravitational wave events like GW190814 ($23.2^{+1.1}_{-1.0} \, M_{\odot}$) and GW200210 ($24.1^{+7.5}_{-4.6} \, M_{\odot}$). To investigate these intriguing phenomena, we adopt the GD approach, which provides critical insights into the formation and characteristics of these astrophysical objects within the framework of standard GR. In our analysis of the SS model, we emphasize how the deformation of the star is influenced by both the decoupling constant and the specific bag function employed in the bag model. This framework necessitates adjustments to the energy and pressure functions based on the premise that quarks are non-interacting and massless. The pressures from the various quark flavors--up, down, and strange--are balanced by the total external bag pressure, resulting in well-defined relationships for matter density and the EOS. We proceed to analyze the spacetime geometries of the system by deriving differential equations that describe the gravitational potentials involved. Through the integration of these equations, we formulate expressions that completely characterize the spacetime geometry of our model. However, the presence of three autonomous equations combined with four unknown variables necessitates the introduction of additional constraints to fully define the system.

Initially, we derive a differential equation that establishes a relationship between the seed density $\epsilon$ and the component $\theta^0_0$. Solving this equation yields the exact solution for the decoupling function $\mathcal{G}(r)$, leading to the formulation of a deformed density constraint. This process results in specific spacetime geometries and defines the components of a new source $\theta_{ij}$. Similarly, we establish a relationship between the seed pressure $P_r$ and the component $\theta^1_1$, deriving another differential equation that allows us to determine $\mathcal{G}(r)$ in this context.

The resulting deformed pressure solution reveals additional spacetime geometries, further characterizing the system and enabling a systematic examination of the impacts of both density and pressure constraints on our model. To construct a realistic compact stellar model, it is essential to connect the inner geometry with the external Schwarzschild spacetime, taking into account additional energy-momentum components that may modify this external structure. By applying junction conditions, we ensure continuity at the boundary, allowing us to derive the necessary relationships and constants that fully define our model. Additionally, we assess the physical viability of our deformed SS models and their significance in astrophysics by analyzing various thermodynamic variables, including energy density, pressure profiles, and mass-radius and mass-moment of inertia relationships. Our findings indicate that the energy density $\epsilon(r)$ decreases from a maximum at the center to a minimum at the surface, with core densities significantly elevated in the GR+MGD scenario. For example, with parameter values of $\epsilon_0 = 0.0003 \, \text{km}^{-2}$ and $\epsilon_s = 0.00024 \, \text{km}^{-2}$, the central density increases notably, highlighting the influence of MGD. We observe that both radial and tangential pressures exhibit a pronounced increase in the central region for the GR+MGD model, indicating a confining effect on the star's core. Specifically, we find that the radial pressure $P_r$ approaches zero at the surface, a critical feature for ensuring stability. The anisotropy parameter $\Delta(r)$ begins at zero at the center and increases towards the outer boundary, suggesting that radial stresses prevail and contribute to the overall stability of the star. Notably, the anisotropy doubles when MGD is considered, underscoring its significant impact on the star's structural integrity.

The mass-radius ($M-R$) and mass-inertia ($M-I$) profiles further illustrate the robustness of our models. For solution \ref{solA}, as the decoupling constant $\beta$ increases from 0 to 0.1 and the bag constant $\mathcal{B}_g$ decreases from 70 $\text{MeV/fm}^3$ to 55 $ \text{MeV/fm}^3$, the maximum mass reaches $M_{\text{max}} = 2.87 \, M_\odot$ with a radius of 11.20 km. In contrast, for $\beta = 0$, the maximum mass is $M_{\text{max}} = 2.48 \, M_\odot$ with a radius of 10.69 km. Similarly, in solution \ref{solB}, as $\beta$ decreases to 0, the maximum mass peaks at $M_{\text{max}} = 2.95 \, M_\odot$ for $\mathcal{B}_g = 55 \, \text{MeV/fm}^3$ with a radius of 11.32 km. These results not only exceed the observed masses of compact stars but also align with findings from recent studies that investigate the properties of nucleonic matter and SSs. For instance, the maximum mass estimates from our models correlate with those observed in gravitational wave events like GW190814, indicating the capability of our models to elucidate astrophysical phenomena. Overall, the interplay of parameters significantly influences the structural properties of these stars, underscoring the relevance of our models in the ongoing exploration of compact objects in the universe.

To evaluate the stability of our anisotropic stellar configuration, we analyze the adiabatic index ($\Gamma$), a crucial parameter that determines whether a stellar structure can withstand perturbations without collapsing. The adiabatic index must exceed a critical threshold, known as the critical adiabatic index ($\Gamma_{cr}$), which increases in the relativistic framework, reflecting additional gravitational influences. Our analysis demonstrates a strong correlation between $\Gamma$ and dynamically stable configurations for both density and pressure constraints. Furthermore, we investigate the relationship between stellar mass ($M$) and central energy density ($\epsilon_c$) using the static stability criterion established by Harrison, Zel'dovich, and Novikov. This criterion states that a negative derivative of mass with respect to central energy density indicates instability, while a positive derivative signifies stability. Our findings reveal that the derivative $\frac{dM}{d\epsilon_0}$ is positive, suggesting that the stellar structures developed are stable. The inclusion of MGD leads to a more significant increase in mass, particularly for the density constraint solution, indicating robustness against small perturbations. Overall, this understanding emphasizes the critical role of both density and pressure in maintaining the structural integrity of anisotropic stellar models, contributing to our comprehension of compact stars, especially in the context of mass gap objects.

 \section*{Acknowledgments}
The author SKM acknowledges that this research work is supported by the TRC Project (Grant No.  BFP/RGP/CBS/24/203). SKM also thankful for continuous support and encouragement from the administration of the University of Nizwa for this research work. AE thanks the National Research Foundation of South Africa for the award of a postdoctoral fellowship.








\begin{thebibliography}{99}

\bibitem{LIGOScientific:2016aoc}
B.~P.~Abbott \textit{et al.} [LIGO Scientific and Virgo],
Phys. Rev. Lett. \textbf{116}, no.6, 061102 (2016)

\bibitem{LIGOScientific:2017ync}
B.~P.~Abbott \textit{et al.} [LIGO Scientific and Virgo],
Astrophys. J. Lett. \textbf{848}, no.2, L12 (2017)


\bibitem{Guo:2023vzz}
L.~Guo and Y.~Niu,
Phys. Rev. C \textbf{110}, no.1, L012801 (2024)

\bibitem{Radice:2017lry}
D.~Radice, A.~Perego, F.~Zappa and S.~Bernuzzi,
Astrophys. J. Lett. \textbf{852}, no.2, L29 (2018)

\bibitem{Burgio:2018yix}
G.~F.~Burgio, A.~Drago, G.~Pagliara, H.~J.~Schulze and J.~B.~Wei,
Astrophys. J. \textbf{860}, no.2, 139 (2018)

\bibitem{Ruiz:2017due}
M.~Ruiz, S.~L.~Shapiro and A.~Tsokaros,
Phys. Rev. D \textbf{97}, no.2, 021501 (2018)

\bibitem{KAGRA:2021duu}
R.~Abbott \textit{et al.} [KAGRA, VIRGO and LIGO Scientific],
Phys. Rev. X \textbf{13}, no.1, 011048 (2023)


\bibitem{Bailyn:1997xt}
C.~D.~Bailyn, R.~K.~Jain, P.~Coppi and J.~A.~Orosz,
Astrophys. J. \textbf{499}, 367 (1998)

\bibitem{Ozel:2010su}
F.~Ozel, D.~Psaltis, R.~Narayan and J.~E.~McClintock,
Astrophys. J. \textbf{725}, 1918-1927 (2010)

\bibitem{Farr:2010tu}
W.~M.~Farr, N.~Sravan, A.~Cantrell, L.~Kreidberg, C.~D.~Bailyn, I.~Mandel and V.~Kalogera,
Astrophys. J. \textbf{741}, 103 (2011)

\bibitem{ArcaSedda:2021zmm}
M.~Arca Sedda,
Astrophys. J. Lett. \textbf{908}, no.2, L38 (2021)

\bibitem{Bhar2023} P. Bhar, A. Errehymy, S. Ray, Eur. Phys. J. C \textbf{83}, 1151 (2023)

\bibitem{Samsing:2019dtb}
J.~Samsing, D.~J.~D'Orazio, K.~Kremer, C.~L.~Rodriguez and A.~Askar,
Phys. Rev. D \textbf{101}, no.12, 123010 (2020)

\bibitem{Ye:2019xvf}
C.~S.~Ye, W.~f.~Fong, K.~Kremer, C.~L.~Rodriguez, S.~Chatterjee, G.~Fragione and F.~A.~Rasio,
Astrophys. J. Lett. \textbf{888}, no.1, L10 (2020)


\bibitem{Fragione:2020aki}
G.~Fragione, A.~Loeb and F.~A.~Rasio,
Astrophys. J. Lett. \textbf{895}, no.1, L15 (2020)

\bibitem{Lu:2020gfh}
W.~Lu, P.~Beniamini and C.~Bonnerot,
Mon. Not. Roy. Astron. Soc. \textbf{500}, no.2, 1817-1832 (2020)

\bibitem{Liu:2020gif}
B.~Liu and D.~Lai,
Mon. Not. Roy. Astron. Soc. \textbf{502}, no.2, 2049-2064 (2021)

\bibitem{LIGOScientific:2020zkf}
R.~Abbott \textit{et al.} [LIGO Scientific and Virgo],
Astrophys. J. Lett. \textbf{896}, no.2, L44 (2020)

\bibitem{Bartos:2023lfu}
I.~Bartos, S.~Rosswog, V.~Gayathri, M.~C.~Miller, D.~Veske and S.~Marka,
[arXiv:2302.10350 [astro-ph.HE]].


\bibitem{Farrow:2019xnc}
N.~Farrow, X.~J.~Zhu and E.~Thrane,
Astrophys. J. \textbf{876}, no.1, 18 (2019)

\bibitem{Kiziltan:2013oja}
B.~Kiziltan, A.~Kottas, M.~De Yoreo and S.~E.~Thorsett,
Astrophys. J. \textbf{778}, 66 (2013)

\bibitem{Valentim:2011vs}
R.~Valentim, E.~Rangel and J.~E.~Horvath,
Mon. Not. Roy. Astron. Soc. \textbf{414}, 1427 (2011)

\bibitem{Zhang:2010qr}
C.~M.~Zhang, J.~Wang, Y.~H.~Zhao, H.~X.~Yin, L.~M.~Song, D.~P.~Menezes, D.~T.~Wickramasinghe, L.~Ferrario and P.~Chardonnet,
Astron. Astrophys. \textbf{527}, A83 (2011)


\bibitem{Ovalle:2017fgl}
J.~Ovalle,
Phys. Rev. D \textbf{95}, no.10, 104019 (2017)

\bibitem{Ovalle:2018gic}
J.~Ovalle,
Phys. Lett. B \textbf{788}, 213-218 (2019)

\bibitem{daRocha:2020jdj}
R.~da Rocha,
Phys. Rev. D \textbf{102}, no.2, 024011 (2020)

\bibitem{Ovalle:2017wqi}
J.~Ovalle, R.~Casadio, R.~da Rocha and A.~Sotomayor,
Eur. Phys. J. C \textbf{78}, no.2, 122 (2018)

\bibitem{Ovalle:2018umz}
J.~Ovalle, R.~Casadio, R.~d.~Rocha, A.~Sotomayor and Z.~Stuchlik,
Eur. Phys. J. C \textbf{78}, no.11, 960 (2018)

\bibitem{Ovalle:2019lbs}
J.~Ovalle, C.~Posada and Z.~Stuchl\'\i{}k,
Class. Quant. Grav. \textbf{36}, no.20, 205010 (2019)

\bibitem{Casadio:2019usg}
R.~Casadio, E.~Contreras, J.~Ovalle, A.~Sotomayor and Z.~Stuchlick,
Eur. Phys. J. C \textbf{79}, no.10, 826 (2019)

\bibitem{Zubair:2020lna}
M.~Zubair and H.~Azmat,
Annals Phys. \textbf{420}, 168248 (2020)

\bibitem{Ovalle:2020kpd}
J.~Ovalle, R.~Casadio, E.~Contreras and A.~Sotomayor,
Phys. Dark Univ. \textbf{31}, 100744 (2021)

\bibitem{Ovalle:2021jzf}
J.~Ovalle, E.~Contreras and Z.~Stuchlik,
Phys. Rev. D \textbf{103}, no.8, 084016 (2021)

\bibitem{Contreras:2021yxe}
E.~Contreras, J.~Ovalle and R.~Casadio,
Phys. Rev. D \textbf{103}, no.4, 044020 (2021)

\bibitem{Maurya:2024ylr}
S.~K.~Maurya, A.~Errehymy, K.~Newton Singh, A.~Aziz, S.~Hansraj and S.~Ray,
Astrophys. J. \textbf{972}, no.2, 175 (2024)

\bibitem{Maurya:2022cyv}
S.~K.~Maurya, A.~Errehymy, R.~Nag and M.~Daoud,
Fortsch. Phys. \textbf{70}, no.5, 2200041 (2022)

\bibitem{Maurya:2022uqu}
S.~K.~Maurya, M.~Govender, K.~N.~Singh and R.~Nag,
Eur. Phys. J. C \textbf{82}, no.1, 49 (2022)

\bibitem{Maurya:2022brt}
S.~K.~Maurya, A.~Banerjee, A.~Pradhan and D.~Yadav,
Eur. Phys. J. C \textbf{82}, no.6, 552 (2022)

\bibitem{Sharp-Misner1964} C. W. Misner, D. H. Sharp, Phys. Rev. 136, B571 (1964).

\bibitem{Gokhroo1994}M.K. Gokhroo, A.L. Mehra, Gen. Relativ. Gravit.26, 75 (1994)
\bibitem{TOV1} R.C. Tolman,  Phys. Rev., 55, 364 (1939).
\bibitem{TOV2} J.R. Oppenheimer and  G.M. Volkoff, Phys. Rev., 55, 374 (1939).
\bibitem{Harko:2002pxr}
T.~Harko and M.~K.~Mak,
Chin. J. Astron. Astrophys. \textbf{2}, no.3, 248 (2002)

\bibitem{Chodos:1974}A. Chodos, R.L.Jaffe, K. Johnson, C.B.Thorn,\& V.F. Weisskopf, Phys. Rev. D 9, 3471 (1974)

\bibitem{Ovalle2017} J. Ovalle, Phys. Rev. D, 95, 104019 (2017)

 \bibitem{OvalleEPJC2018}  J. Ovalle, R. Casadio, R. da Rocha, A. Sotomayor, Eur. Phys. J. C, 78, 122 (2018)

\bibitem{sharif2} M. Sharif, A. Majid,  Phys. Dark Univ. 30, 100610 (2020)

\bibitem{Maurya:2023muz}
S.~K.~Maurya, K.~N.~Singh, M.~Govender, G.~Mustafa and S.~Ray,
Astrophys. J. Suppl. \textbf{269}, no.2, 35 (2023)

\bibitem{Maurya:2021zvb}
S.~K.~Maurya, K.~N.~Singh, M.~Govender and S.~Hansraj,
Astrophys. J. \textbf{925}, no.2, 208 (2022)

\bibitem{Demorest:2010bx}
P.~Demorest, T.~Pennucci, S.~Ransom, M.~Roberts and J.~Hessels,
Nature \textbf{467}, 1081-1083 (2010)

\bibitem{Romani:2022jhd}
R.~W.~Romani, D.~Kandel, A.~V.~Filippenko, T.~G.~Brink and W.~Zheng,
Astrophys. J. Lett. \textbf{934}, no.2, L17 (2022)

\bibitem{KAGRA:2021vkt}
R.~Abbott \textit{et al.} [KAGRA, VIRGO and LIGO Scientific],
Phys. Rev. X \textbf{13}, no.4, 041039 (2023)

\bibitem{MI} M. Bejger and P. Haensel, 
Astron. Astrophys. \textbf{396}, 3 (2002)

\bibitem{Astashenok:2021peo}
A.~V.~Astashenok, S.~Capozziello, S.~D.~Odintsov and V.~K.~Oikonomou,
Phys. Lett. B \textbf{816}, 136222 (2021)

\bibitem{Astashenok:2020qds}
A.~V.~Astashenok, S.~Capozziello, S.~D.~Odintsov and V.~K.~Oikonomou,
Phys. Lett. B \textbf{811}, 135910 (2020)

\bibitem{Astashenok:2020cqq}
A.~V.~Astashenok and S.~D.~Odintsov,
Mon. Not. Roy. Astron. Soc. \textbf{498}, no.3, 3616-3623 (2020)

\bibitem{Astashenok:2020cfv}
A.~V.~Astashenok and S.~D.~Odintsov,
Mon. Not. Roy. Astron. Soc. \textbf{493}, no.1, 78-86 (2020)

\bibitem{Astashenok:2013vza}
A.~V.~Astashenok, S.~Capozziello and S.~D.~Odintsov,
JCAP \textbf{12}, 040 (2013)

\bibitem{Astashenok:2014nua}
A.~V.~Astashenok, S.~Capozziello and S.~D.~Odintsov,
JCAP \textbf{01}, 001 (2015)

\bibitem{Astashenok:2017dpo}
A.~V.~Astashenok, S.~D.~Odintsov and A.~de la Cruz-Dombriz,
Class. Quant. Grav. \textbf{34}, no.20, 205008 (2017)

\bibitem{Odintsov:2023ypt}
S.~D.~Odintsov and V.~K.~Oikonomou,
Phys. Rev. D \textbf{107}, no.10, 104039 (2023)
doi:10.1103/PhysRevD.107.104039

\bibitem{Nashed:2021sji}
G.~G.~L.~Nashed and S.~Capozziello,
Eur. Phys. J. C \textbf{81}, no.5, 481 (2021)

\bibitem[\protect\citeauthoryear{Bondi}{1992}]{bondi1992anisotropic}  Bondi, H. 1992, MNRAS, 259, 365.

\bibitem[\protect\citeauthoryear{ Heintzmann \& Heintzmann}{1975}]{heintzmann1975neutron}  Heintzmann, H., \& Heintzmann, W. 1975, A\&A, 38, 51

\bibitem[\protect\citeauthoryear{Chan et al.}{1993}]{chan1993dynamical}  Chan, R.,   Herrera, L.,  \& Santos, N. O. 1993, MNRAS, 265, 533

\bibitem[\protect\citeauthoryear{Chan et al.}{1992}]{chan1992dynamical}  Chan, R.,  Herrera, L.,   \& Santos, N. O. 1992, CQGra, 9, 133

\bibitem{Moustakidis:2016ndw}
C.~C.~Moustakidis,
Gen. Rel. Grav. \textbf{49}, no.5, 68 (2017)

\bibitem{Koliogiannis:2018hoh}
P.~S.~Koliogiannis and C.~C.~Moustakidis,
Astrophys. Space Sci. \textbf{364}, no.3, 52 (2019)

\bibitem{ZHN1} B.K. Harrison, K.S. Thorne, M. Wakano, J.A. Wheeler, Gravitational Theory and Gravitational Collapse, University of Chicago Press (1965)

\bibitem{ZHN2} Ya.B. Zeldovich, I.D. Novikov, Relativistic Astrophysics Stars and Relativity, Vol. 1, University of Chicago Press (1971)

\end{thebibliography}
\end{document}